\documentclass{aa}  
\usepackage{graphicx}
\usepackage{float}
\usepackage{longtable}
\usepackage{txfonts}
\begin{document}
   \title{Chemical composition of A and F dwarf members of the Coma Berenices open cluster\thanks{Based on observations at the Observatoire de Haute-Provence (France).}}


   \author{M. Gebran\inst{1}, R. Monier\inst{1}\thanks{Present affiliation: Laboratoire Universitaire d'Astrophysique de Nice, UMR 6525, Universit\'e de Nice - Sophia Antipolis, Parc Valrose, 06108 Nice Cedex 2, France} \and O. Richard\inst{1} }

   \offprints{M. Gebran}

   \institute{$^1$ Groupe de Recherche en Astronomie et Astrophysique du Languedoc,UMR 5024, Universit\'e Montpellier II, Place Eug\`ene Bataillon, 34095 Montpellier, France. \\
    \\   
   \email{gebran@graal.univ-montp2.fr}\\
   \email{Richard.Monier@unice.fr}\\
   \email{richard@graal.univ-montp2.fr}}

   \date{Received ; accepted }

 
  \abstract
   {}
   {Abundances of 18 chemical elements have been derived for 11 A (normal and Am) and 11 F dwarfs members of the Coma Berenices open cluster in order to set constraints on evolutionary models including transport processes (radiative and turbulent diffusion) calculated with the Montr\'eal code.}
   {A spectral synthesis iterative procedure has been applied to derive the abundances from selected high quality lines in high resolution high signal-to-noise \'echelle spectra obtained with ELODIE at the Observatoire de Haute Provence.} 
     {The chemical pattern found for the A and F dwarfs in Coma Berenices is reminiscent of that found in the Hyades and the UMa moving group. In graphs representing the abundances [X/H] versus the effective temperature, the A stars often display abundances much more scattered around their mean values than the F stars do. Large star-to-star variations are detected for A stars in their abundances of C, O, Na, Sc, Ti, Mn, Fe, Ni, Sr, Y, Zr and Ba which we interpret as evidence of transport processes competing with radiative diffusion.\\
     The abundances of Mn, Ni, Sr and Ba are strongly correlated with that of iron for A and Am stars. In contrast the ratios [C/Fe] and [O/Fe] appear to be anticorrelated with [Fe/H] as found earlier for field A dwarfs. All Am stars in Coma Berenices are deficient in C and O and overabundant in elements heavier than Fe but not all are deficient in calcium and/or scandium. The F stars have solar abundances for almost all elements except for Mg, Si, V and Ba.\\
     The derived abundances patterns, [X/H] versus atomic number, for the slow rotator HD108642 (A2m) and the moderately fast rotator HD106887 (A4m) were compared to the predictions of self consistent evolutionary model codes including radiative and different amounts of turbulent diffusion. None of the models reproduces entirely the overall shape of the abundance pattern.}
   {While part of the discrepancies between derived and predicted abundances may be accounted for by non-LTE effects, the inclusion of competing processes such as rotational mixing in the radiative zones of these stars seems necessary to improve the agreement between observed and predicted abundance patterns.}


   \maketitle


\section{Introduction}

Abundance determinations of A and F dwarfs in open clusters and moving groups of known properties aim at 
elucidating the mecanisms of mixing at play in the interiors of these main-sequence stars.
In two previous papers, abundance determinations were presented for 11 chemical
elements in the Hyades (age about 787 Myrs, \cite{varenne-monier-99}) and in the Ursa Major moving
group (age about 500 Myr, \cite{monier2005}). Specifically, 19 A and 29 F dwarfs
members of the Hyades
and 12 A and 10 F dwarf bona-fide and probable
members of the Ursa Major moving group were analysed. The motivation of this
study is to report on new abundance determinations of 18 chemical elements in 11
A and 11 F dwarfs in the Coma Berenices open cluster (age about 447 Myrs). The spectroscopy presented here is part of an ongoing
observational program
of A and F dwarfs in galactic open clusters of various ages whose aims are
twofold. First, we wish to improve
our knowledge of the chemical composition of A dwarfs (normal A stars and Am
stars) which is still poor in particular for normal A stars. Second, we intend 
to use the derived abundances to set constraints on self-consistent evolutionary models of
these objects including various particle transport processes. 
Indeed stars in open clusters originate from the same interstellar material 
  ({\it ie} they have the same age and the same {\it
 initial} chemical composition) and as such are very useful to test
 the predictions of evolutionary models.
\\
Few studies have addressed the chemical composition of the A and F dwarfs in the 
Coma Berenices open cluster.
High quality high resolution spectroscopy is feasible for the brightest members because of the
proximity (d $\simeq 80-90 $pc) of the cluster. Abundance analyses of the Coma
Berenices dwarfs
have mostly focused on
F and G dwarfs whose low apparent rotational velocities facilitate the abundance
determination. 
For concision, we have collected previous 
 abundance determinations
of the A, F and G dwarfs in Coma Berenices in Table \ref{previous-determinations} where the bibliographical references, the
number of stars analysed, their spectral types and the
 investigated chemical elements are collected.
   Lithium abundances have been determined for several F (and G) dwarfs in
  Coma Berenices by  \cite{Boesgaard1987}, \cite{jeffries1999} and \cite{soderblometal990}. Carbon and iron abundances were derived for 14 F dwarfs by \cite{FB92}. Carbon, oxygen, silicon, calcium, iron, barium, magnesium, scandium, chromium, nickel, lithium,
 aluminium, sulfur and europium abundances 
 have been determined for only a few normal A and Am stars (fewer than 7 stars) by Savanov (1996), Hui-Bon-Hoa et al. (1997), Hui-Bon-Hoa \& Alecian (1998) and \cite{burkhart-coupry-2000}. The abundance determinations for A stars focuse mainly on the chemically peculiar Am stars.
 \cite{savanov96} found significant star-to-star differences in abundances among the
 Am stars of Coma Ber for a given chemical element. 
 In contrast, little attention has been paid to the "normal" A
 stars of the cluster. 
 The chemical composition of normal A dwarfs (field and cluster stars) remains generally poorly known
  as too few 
 objects have been
 analysed so far, mainly because of their high rotational velocities. 
 Significant abundance differences have been found among the few field 
"normal" A stars analysed so far (\cite{1986A&A...163..333H}; \cite{1986PASP...98..927L}; Lemke 1998,1990; Hill \& Landstreet 1993; Hill 1995; Rentzsch-Holm 1997; Varenne 1999). 
 Varenne \& Monier (1999) also found significant star-to-star variations of the
 abundances of O, Na, Ni, Y and Ba for the normal A and the Am stars in the
 Hyades whereas the F dwarfs display much less scatter. Similarly, Monier (2005) found
 star-to-star variations in [Fe/H], [Ni/H] and [Si/H] much larger for the A
 dwarfs than for the F dwarfs in the Ursa Major group.
 \\
 The main thrust of this paper is to report on the abundances of 18 chemical
 elements (C, O, Na, Mg, Si, Ca, Sc, Ti, V, Cr, Mn, Fe, Co, Ni, Sr, Y, Zr and Ba)
 for F and A dwarfs in Coma Berenices. These abundances have been 
   compared to published predictions of recent self consistent models
  including transport processes at ages close to that of Coma Berenices (Turcotte et al. 1998,
  Richer et al. 2000, Richard et al. 2002) or to new models calculated with the Montr\'eal code 
  (Richard et al. 2002).  The abundances were derived by
  synthesizing  carefully selected lines in high quality high
 resolution spectra of 11 A and Am stars and 11 F stars of the cluster.
The selection of the sample of stars observed, the observations and the data reduction are described in Section 2.
The determination of the fundamental parameters, the construction of the linelist and the spectrum synthesis are discussed in Section
3. The behaviour of the abundances of individual chemical elements in the A and F dwarfs of Coma Berenices versus effective temperature and the abundance of iron are described in Section 4. The astrophysical implications of our findings and the detailed comparison of the found abundance patterns for 2 Am stars with recent self-consistent evolutionary models including transport processes are presented in Section 5. Conclusions are given in Section 6.\\

\begin{table*}[htbp]
\caption{Previous abundance determinations for the Coma Ber A, F and G dwarfs.}
\label{previous-determinations}
\begin{center}
\begin{tabular}{c c c c}
\hline
 & & \\
Reference & Stars studied & Chemical Elements \\
\hline
 & & \\
 Savanov (1996)             & 13 A-Am and F-Fm  &  C,O,Si,Ca,Fe,Ba  \\
 Hui-Bon-Hoa et al. (1997)  &  2 A-Am  & Mg,Ca,Sc,Cr,Fe,Ni        \\
 Hui-Bon-Hoa \& Alecian (1998) & 4 A-Am & Mg,Ca,Sc,Cr,Fe,Ni \\
 Burkhart \& Coupry (2000)  & 7 A-Am &  Li,Al,Si,S,Fe,Ni,Eu\\
  \hline
 Boesgaard (1987)         & 22 A and F   &   Li  \\
Friel \& Boesgaard (1992) & 14 F    & Fe,C \\
Jeffries (1999)           & 15 F,G,K  & Li \\
\hline
Cayrel et al. (1988)      & 4 G   &  Fe \\
 Soderblom et al. (1990)  & 28 G  & Li \\
\hline
\end{tabular}
\end{center}
\end{table*} 

\section{Program stars, observations and data reduction}
\label{reduction}
Our observing sample consists of all A and F stars members of the Coma Berenices cluster brighter than V = 8.6 mag retrieved from Trumpler's (1938) list.
This magnitude corresponds to the latest F dwarfs (F8V) which required the longest achievable exposure times (about 1 hour 15 minutes).
Eleven A ("normal" and Am stars) stars and 11 F stars members of the Coma Berenices
cluster were observed using two spectrographs at Observatoire de Haute Provence (OHP): the ELODIE \'echelle spectrograph and the monoorder AURELIE spectrograph.
AURELIE is a monoorder spectrograph (Gillet et al. 1994) placed at the coud\'e focus of the 152 cm telescope. AURELIE is free of scattered light and has been used to calibrate putative effects of scattered light on the line profiles in the blue region obtained with ELODIE.
ELODIE is a fiber-fed cross-dispersed \'echelle spectrograph attached to 193 cm at OHP (Baranne et al. 1996). It records in a single exposure a spectrum extending from 3850 \AA\ to 6811 \AA\ at a resolving power of about 42000 on a relatively small CCD (1024X1024). 
The observing dates, exposure times and Signal to Noise ratios achieved for each star are collected in Table \ref{obs}, the first part being dedicated to ELODIE spectra while the second describes the AURELIE spectra obtained in three 70 \AA\ wide regions centered around 4505 \AA, 5080 \AA, 5530 \AA \ and 6160 \AA.
 
The fundamental data of these stars are collected in Table \ref{etoiles-coma-vitesses}.
The Trumpler and Henry Draper identifications appear in columns 1 and 2, the spectral types retrieved from SIMBAD in column 3, the apparent magnitudes in column 4. Columns 5 to 8 display the effective temperatures and surface gravities adopted for the analysis and the rotational velocities and microturbulent velocities  derived in our analysis (see section \ref{takeda}). Comments about binarity and pulsation appear in the last column. Note that 2 stars, HD 107655 and HD 106999, may not be members of the cluster (Bounatiro 1993). Although we derived abundances for these stars too, their data do not appear in the figures.
 There are no very rapid rotators in this cluster, the apparent rotational velocities range from 9 km$\cdot$s$^{-1}$ to 102 km$\cdot$s$^{-1}$.\\
Inspection of the Catalogue of Double and Multiple stars (CCDM, Dommanget \& Nys 1995) reveals that only one star is a binary system. HD 106887 is the primary star in a double system and has a much fainter companion (V = 9.8) located nearby at about 8.6 arcsec, whose spectral type is unknown. It is unlikely that the light of the companion may have contributed significantly to the observed spectra of HD 106887. 
\\
With AURELIE, four spectral regions centered
 on $\lambda$4505 \AA\ (region 1), $\lambda$6160\AA\ (region 2), $\lambda$5080\AA\ (region 3) and 
 $\lambda$5530\AA\ (region 4) have been 
observed as they include several lines having accurate oscillator strengths for the chemical elements we study. Region 1 contains the Mg 2 triplet at $\lambda$4481 \AA\
and several clean unblended and well separated Fe 2 lines. Regions 2 to 4 are those observed by Edvardsson et al. (1993) in their spectroscopic survey of F
 dwarfs in the galactic disk. 
 For the brightest stars (V $\leq$ 6), grating 5 working at its second order (for region 1) and 
 grating 1 (for region 2) yielded resolving powers equal to 
 60000 and 65000 respectively. For the
 fainter stars (V $>$ 6), grating 7 was used, resulting in resolving
 powers of 36000 and 28000 (for regions 1 and 2 respectively). In region 3, 
 all stars have been observed with grating 7 (R = 31000) except for Procyon.  
 According to the V magnitude of the stars and weather conditions, 
 exposure times between 1h30 and 3 hours were necessary to secure signal to noise ratios around
 200. The IRAF software was used to reduce all spectra following the standard procedure 
 (mean offset removal, division by the mean flat field, wavelength
calibration and continuum normalization). 
\\
\begin{table*}
\caption{Data on the programme stars. Spectral type are taken from 
SIMBAD and WEBDA online database. $T_{\rm{eff}}$ and $\log g$ are those 
determined by UVBYBETA code. $v_{e}\sin i$  and $\xi_{t}$ are determined 
as explained in sect. \ref{takeda}. References (a) and (b) are 
\cite{1993A&AS..100..531B} and \cite{1994A&AS..106...21R} respectively.}
\label{etoiles-coma-vitesses}
\centering
\begin{tabular}{lcccccccc}
\hline
  TR&HD       & Type     & $m_{v}$    &$T_{\rm{eff}}$          &  $\log 
g$     &       $v_{e}\sin i$  & $\xi_{t}$ & Remarques \\
    &       &         &         &(K)                  &          &    
km$\cdot$s$^{-1}$             &km$\cdot$s$^{-1}$       &  \\ \hline
19   &   HD106103  &F5V         &8.09     &    6707           &  4.45    &      19.7        &      1.4     &           \\
     &     HD106293  &F5V         &8.09     &    6545           &4.34    &      47        &      1.4     &           \\
36   &     HD106691  &F5IV        &8.08     &    6713           &4.43    &      37        &      1.6     &           \\
49   &     HD106946  &F2V         &7.87     &    6892           &4.30    &      62        &      1.9     &           \\
86   &     HD107611  &F6V         &8.50     &    6491           &4.57    &      22        &      1.4     &           \\
     &     HD109530  &F2V         &7.30     &    6497           &3.85    &      67        &      2.1     &           \\
101  &     HD107877  &F6          &8.35     &    6598           &4.54    &      29.5        &      1.25    &           \\
114  &     HD108154  &F5          &8.56     &    6497           &4.54    &      19        &      1.15    &           \\
118  &     HD108226  &F5          &8.34     &    6530           &4.45    &      19        &      1.05    &           \\
162  &     HD108976  &F6 V        &8.54     &    6413           &4.49    &      20        &      1.1     &           \\
     &     HD109069  & F0 V        &7.55     &    6864          &4.06     &     89        &     2.2      &           \\  \hline \\ \hline
107  &     HD107966  &A3V/A3IV        &5.18     &   8541        &3.82      &   51          &  2.9         &           \\
130  &     HD108382  &A4V/A3IV        &4.96     &   8317        &3.92      &   75.5        &  3.2         &           \\
47   &     HD106887  &A4m         &5.71     &   8291           &4.20    &   82          &  3.8         &           \\
86   &     HD107655  &A0V         &6.18     &   9675           &  4.10  &   45          &  2.0       &$\notin$ Coma (a)       \\
62   &     HD107168  &A8m/kA5hA5mF0   &6.24     &   8283           & 4.20        &   14.3        &  4.0         &           \\
183  &     HD109307  &A4Vm/A3IV-V     &6.26     &   8396    &  4.10&   14.5        &  3.3      &            \\
144  &     HD108642  &A2m/kA2hA7mA7   &6.54     &   8079           &4.06      &   9.2         &  3.7         &           \\     
68   &     HD107276  &Am/Ka5mA7        &6.63     &   8000           &4.00      &   102         &  2.8         &           \\
139  &     HD108486  &AmkA3hA5mA7     &6.67     &   8148           & 4.11      &   37          &  3.0         &           \\
52   &     HD106999  &Am          &7.46     &    8148           &    4.09    &      44        &      3.0       &       $\notin$ Coma (a)       \\
82   &     HD107513  &Am/kA7hF0mF0    &7.38     &    7279           &4.02    &      62        &      3.0       &       $\delta$ Scuti (b) 
\\   \hline
 \hskip0.1 cm
\end{tabular}
 \end{table*}
At the end of an observing night, the ELODIE spectra are provided to the observer fully reduced by the INTER-TACOS (INTERpreter for the Treatment, the Analysis and the COrrelation of Spectra) pipeline developed by D. Queloz and L. Weber (Baranne et al. 1996). The reduction is actually included in the spectrograph data flow and provides fully flat-fielded and wavelength calibrated spectra directly at the end of each exposure. However, we have chosen to perform our own reduction of the ELODIE spectra. Indeed, ELODIE was primarily designed to provide accurate radial velocity measurements. One source of concern when deriving abundances is to properly correct the spectrum for scattered light, especially in the blue region. The background in a stellar ELODIE exposure can be estimated by measuring the flux in the inter-orders. In INTER-TACOS, this background is removed using a two dimensional polynomial fit with a typical 5 \% error which peaks in the middle of the orders (see fig. 11 in Baranne et al. 1996). Erspamer \& North (2002) have devised a reduction procedure using a set of IRAF functions which they apply to the raw image in order to provide an improved correction for scattered light compared to INTER-TACOS. We have also chosen to perform our own reduction of the ELODIE spectra using IRAF (Image Reduction and Analysis Facility, Tody 1993) routines. Although it follows Erspamer \& North's (2002) procedure, our reduction slightly differs from their method. The steps are as follows:
\begin{itemize}
\item[1-] Averaging the several offsets and flat-fields taken throughout the night, using $\bf{zerocombine}$ and $\bf{flatcombine}$.
\item[2-] Removal of the mean offset from all images and bad pixels correction using $\bf{ccdproc}$ .
\item[3-] Finding and centering the orders using the flat-field image using $\bf{apfind}$ and $\bf{apcenter}$. 
\item[4-] Removal of the scattered light ($\bf{apscatter}$). Scattered light fills in the line profiles, making the lines shallower and thus leading to underabundances if not taken into account. The scattered light is estimated (as explained in sect 3.2 of Erspamer \& North 2002) and substracted from the original image. An example of the effect of this improved removal of the scattered light is shown in Figure \ref{diff-tacos-iraf}, where the corrected profiles of the FeII lines at 4520.224 \AA \ and 4522.634 \AA \ are compared to the INTER-TACOS profiles. In this case, ignoring the scattered light would lead to underestimation of the iron abundance deduced from these 2 lines by 0.07 and 0.08 dex respectively. The effect is more pronounced, about 0.14 dex, for the MgII triplet at 4481 \AA.
\item[5-] Extraction of the images using $\bf{apsum}$. The averaged flat-field is used to determine the shape of the orders which is used later as a reference for the extraction of the images. The extraction method uses Horne's (1986) algorithm.
\item[6-] Calibration of the thorium image using $\bf{apsum}$, $\bf{ecidentify }$ and $\bf{ecreidentify}$.  
\item[7-] Division by the extracted flat-field using $\bf{sarith}$. 
\item[8-] Wavelength calibration of the spectra using the thorium spectra using $\bf{dispcor}$. 
\item[9-] Normalization to the continuum using $\bf{continuum}$. To ensure we correctly located regions free of lines (when available) in each order, we have computed synthetic spectra using the code SYNSPEC48 (Hubeny \& Lanz 1992) assuming a solar metallicity for the various temperatures and surface gravities of our stars. The spectrum was then rectified to this local continuum.
\item[10-] Merging the 67 normalized orders using $\bf{scombine}$. 
The last three orders do not overlap and thus could not be merged. For the abundance analysis, we have discarded lines located in the overlapping region of two successive orders.
\item[11-] Radial velocity determination using $\bf{fxcor}$: the final merged spectrum is cross-correlated with differents masks of spectral types A0V, A5V, A9V and F5V to derive the radial velocity. The merged spectrum is then corrected for the radial velocity found.
\end{itemize}
Figure \ref{comparaison-spectres-observe-vrot} compares the spectral order 21, centered around $\lambda 4500$ \AA, for 4 A and 4 F stars reduced in this manner. The effect of increasing stellar rotation on line profiles is conspicuous. The last plot of Figure \ref{comparaison-spectres-observe-vrot} displays a typical agreement between the observed spectrum of HD 107655 (A0V) and the synthetic spectrum (computed as explained in sect. \ref{abundances-analysis}) that provides the best fit.

\begin{figure}
\centering
\includegraphics[scale=0.4]{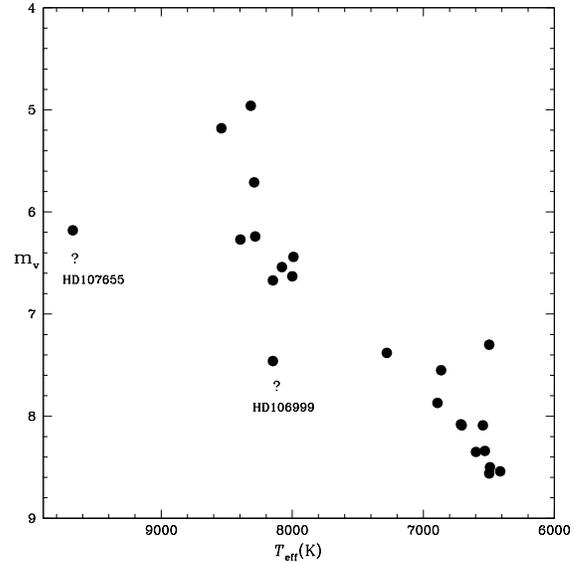}
\caption{Hertzsprung-Russel diagram of the observed stars in Coma Berenices (visual magnitude versus $T_{\rm{eff}}$). The two stars flagged with question marks are probably not members of the cluster.}
\label{Coma-Mv}
 \end{figure}

\begin{table*}
\caption{Observing log of the programme stars of Coma Berenices. The first table describes the ELODIE observations and the second one the AURELIE observations}
\label{obs}
\centering
\begin{tabular}{||c|ccc|c|c||}
\hline  \hline
HD&spectral &$M_{V}$&exposure & S/N &Date\\ 
& type & mag  & time (s)&& \\ \hline
 106103 &   F5V       &  8.09	       & 4500	   &144     &	04/10/04		\\ 
 106293 &   F5V       &  8.09	       &  4500    &132      &	04/10/04	\\  
 106691 &   F5IV      &  8.08	       &4500	   &147     &	04/10/04	 \\	    
 106946 &   F2V       &  7.87	       &  4500    &180      &	04/11/04	\\ 
 107611 &   F6V       &  8.50	       &4500	  &129      &	04/11/04	\\ 
 109530 &   F2V       &  7.30	       &  4500    &185      &	04/12/04	\\ 
 106887 &   A4m       &  5.71	       &  3600    &283      &	04/07/04	\\ 
 106999 &   Am        &  7.46	       &  4500    &177      &	04/10/04	\\ 
 107131 &   A6IV-V    &  6.44	       &  4500    &213      &	04/09/04	\\ 
 107168 &   A8m       &  6.24	       &  4500    &215      &	04/07/04	\\ 
 107276 &   Am        &  6.63	       &  4500    &152      &	04/09/04	\\ 
 107513 &   Am        &  7.38	       &  4500    &184      &	04/10/04		\\ 
 107655 &   A0V       &  6.18	       &  4500    &254      &	04/07/04	\\ 
 107966 &   A3V       &  5.18	       &  3600    &336      &	04/07/04	\\ 
 108382 &   A4V       &  4.96	       &  3600    &337      &	04/07/04	\\ 
 108486 &   Am        &  6.67	       &  4500    &121      &	04/09/04	\\ 
 108642 &   A2m       &  6.54	       &  4500    &170      &	04/09/04	\\ 
 108651 &   A0p       &  6.65	       &  4500    &450      &	04/11/04	\\ 
 109307 &   A4Vm      &  6.26	       &  3600    &114      &	04/08/04	\\ \hline \hline
\end{tabular}
\begin{tabular}{||c|cc|c|c||c|c|c||}
\hline  \hline
HD&spectral &$M_{V}$& grating 5: $\lambda_{c}$=4505 \AA &Date &grating 7: $\lambda_{c}$=6160 \AA& $\lambda_{c}$=5080 \AA &$\lambda_{c}$=5530 \AA\\ 
& type & mag  & exp.time(s)& &exp.time(mn)&exp.time(mn)&exp.time(mn)\\ \hline
107168 &   A8m        &6.24	     &800	  & 03/14/04&	 75     & 104	       &120 (grating 1) \\
107513 &   Am	     &  7.38	     &2000	  & 03/15/04&	  120   & 120	       &120	   \\
106103 &   F5V       &  8.09	     &4500	  & 03/15/04&	  60    &	       &	   \\
107655 &   A0V       &  6.18	     &800	  & 03/15/04&	  135   &85	       &90	   \\
106887 &   A4m       &  5.71	     &  600	  & 03/15/04&	        &	       &60	   \\
107131 &   A6IV-V    &  6.44	     &  800	  & 03/15/04&	  110   &	       &	   \\
107276 &   Am	     &  6.63	     &1000	  & 03/15/04&	  90    & 120	       &75	   \\
107966 &   A3V       &  5.18	     &500	  & 03/15/04&	  45    &  105         &40	   \\
108382 &   A4V       &  4.96	     &500	  & 03/15/04&	  60    &  74	       &	   \\
108486 &   Am	     &  6.67	     &1000	  & 03/15/04&	        & 180	       &90	   \\
108642 &   A2m       &  6.54	     & 1000	  & 03/15/04&	  90    & 120	       &	   \\
108651 &   A0p       &  6.65	     & 1000	  & 03/15/04&	  105   & 150	       &	   \\
109307 &   A4Vm      &  6.26	     &  800	  & 03/15/04&	 120    & 120 (grating 1)	    &80  \\
106293 &   F5V       &  8.09	     &  4500	  & 03/16/04&	  130   &	     &          \\
106691 &   F5IV      &  8.08	     &4500	  & 03/16/04&	  120   & 150	     &120       \\
109069 &   F0V       &  7.55	     &  3000	  & 03/16/04&	  100   &150	     &110        \\	  
107877 &   F6	     &  8.35	     &4500	  & 03/16/04&	  180   &165	     &           \\ 
107611 &   F6V       &  8.50	     &4500	  & 03/16/04&	  90    &	     &          \\
108154 &   F5	     &  8.56	     &  4500	  & 03/17/04&	  180   &	     &165        \\ 
108226 &   F5	     &  8.34	     &  4500	  & 03/17/04&	  90    &180	     &           \\ 
108976 &   F6V       &  8.54	     &  4500	  & 03/17/04&	  150   &	     &140        \\ 
109530 &   F2V       &  7.30	     & 3000	  & 03/17/04&	  100   &120	     &110       \\ \hline 
\end{tabular}
\end{table*}

\begin{figure}
\centering
\includegraphics[width=8cm]{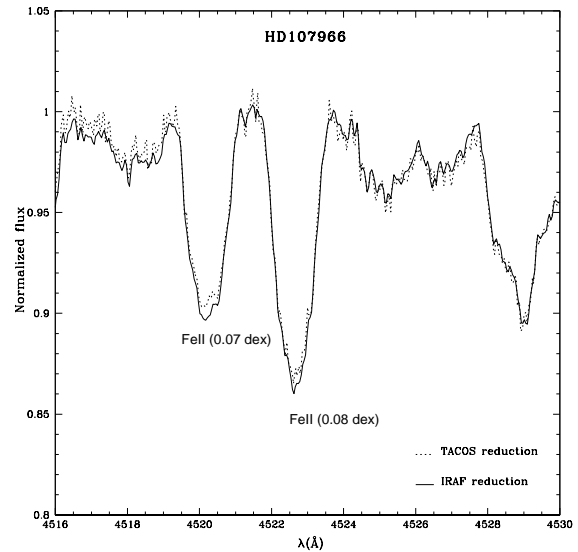}
\caption{Effect of a proper removal of scattered light. The corrected profiles (IRAF reduction, thick lines) are deeper than the INTER-TACOS ones (dashed lines). The abundances deduced from the corrected profiles are about 0.08 dex larger.}
\label{diff-tacos-iraf}
 \end{figure}
 
 \begin{figure*}
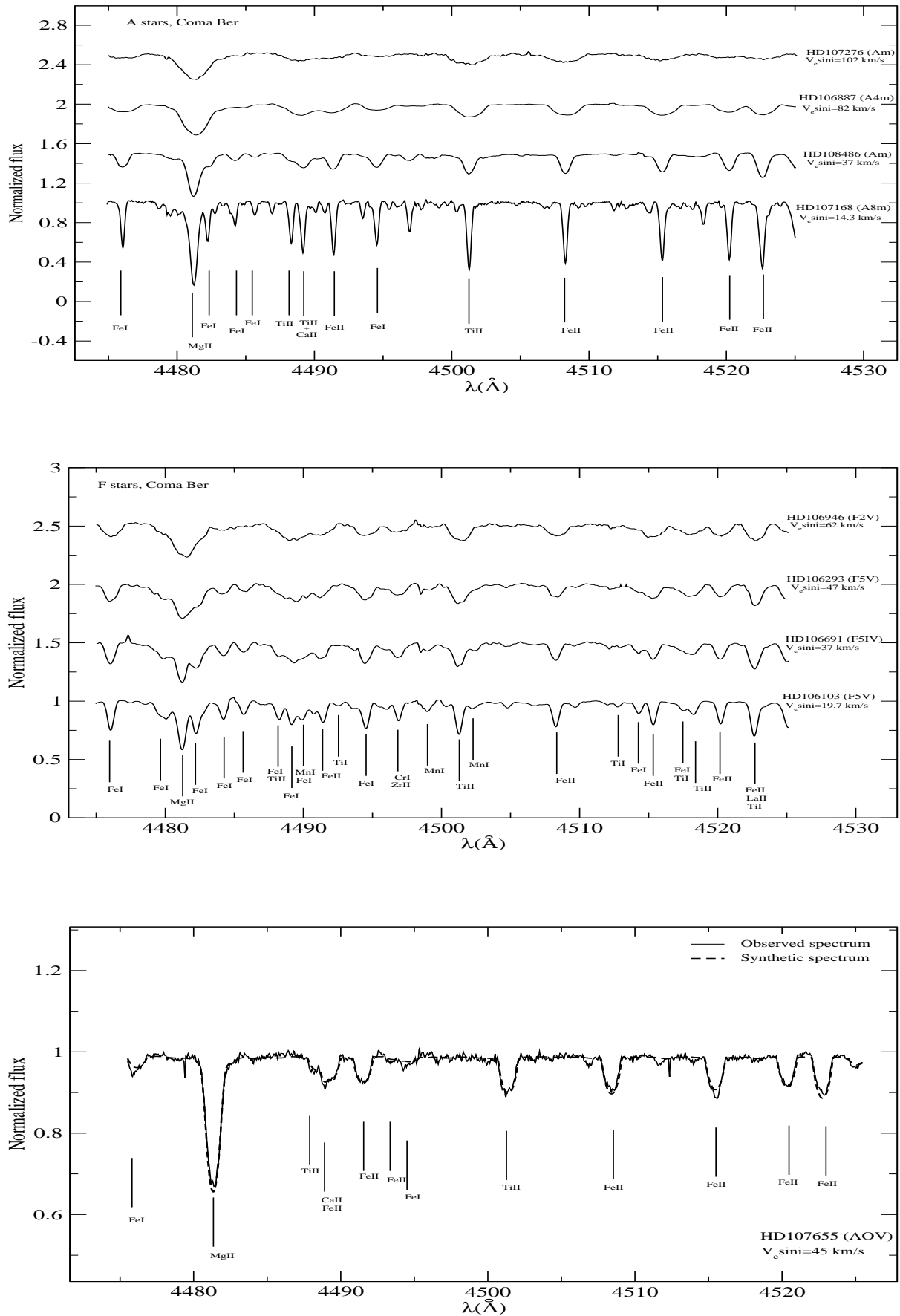

\centering
\vskip1.2cm
\includegraphics[width=16cm,height=7cm]{8807fig3.eps} \\ \vskip1.15cm
\includegraphics[width=16cm,height=7cm]{8807fig4.eps} \\ \vskip1.28cm
\includegraphics[width=16cm,height=7cm]{8807fig5.eps} 
\caption{Selected observed spectra of A (top) and F (middle) dwarf stars members of Coma Berenices open cluster. Spectra are arbitrarily shifted vertically by 0.5, 1 and 1.5 unit of normalized flux. The smearing out of the spectra by rotation is noticeable. The bottom figure displays the final synthetic spectrum (dashed thick line) superimposed on the observed one (thin line) for the star HD107655 (A0V). Identifications for the most intense lines in each region are provided.}
\label{comparaison-spectres-observe-vrot}
 \end{figure*}
	  
\section{Abundance analysis}
\label{abundances-analysis}
The abundances of 18 chemical elements have been derived by iteratively adjusting synthetic spectra to the observed normalized spectra and minimizing the chisquare of the models to the observations. 
Spectrum synthesis is the most appropriate method as our stars have apparent rotational velocities ranging from a 9 to 102 km$\cdot$s$^{-1}$. Specifically, 
synthetic spectra were computed assuming LTE using Takeda's (1995) iterative
procedure and double-checked using Hubeny \& Lanz's (1992) SYNSPEC48 code. This version of SYNSPEC calculates lines for elements heavier than Zn up to Z=99.

\subsection{Atmospheric parameters and model atmospheres}

The effective temperatures ($T_{\rm{eff}}$) and surface gravities 
(log{\it g}) of the stars have been determined using the \cite{1993A&A...268..653N} UVBYBETA calibration of the Str\"omgren photomery indices 
{\it uvby} in terms of $T_{\rm{eff}}$ and log{\it g}. The found effective temperatures and surface gravities are collected in Table \ref{etoiles-coma-vitesses}. The errors on $T_{\rm{eff}}$ 
and log{\it g} are estimated to be $\pm$ 125 K and $\pm$ 0.2 dex respectively.
Model atmospheres were then calculated using Kurucz's ATLAS9 code (Kurucz, 1979),
assuming a plane parallel geometry, a gas in hydrostatic and radiative
equilibrium and LTE. The ATLAS9 model atmospheres contain 64 layers with a regular increase in  $\log \tau_{Ross}$ = 0.125 and they have been calculated assuming Grevesse \& Sauval (1998) solar chemical composition. This ATLAS9 version uses the new opacity distribution function (ODFs) (\cite{2003IAUS..210P.A20C})
calculated for this solar chemical composition.
The line opacity calculation uses the 58 million linelist compiled by
Kurucz (Kurucz 1992a, 1992b).
Convection in ATLAS9 relies on the mixing length theory (MLT). Specifically, we have adopted Smalley's prescriptions (\cite{2004IAUS..224..131S}) for the values of the ratio of the mixing length to the pressure scale height ($\alpha=\frac{L}{H_{P}}$) and also for the microturbulent velocity (constant with depth).
        
\subsubsection{The linelist}

The linelist was constructed from Kurucz's gfall.dat\footnote{http://kurucz.harvard.edu/LINELISTS/GFALL/} list, from which we selected lines between 3000 and 7000 \AA. However the data in this list have been modified and complemented in different manners.
For the lines which we expected to contribute significantly to the absorption, we have carefully checked the wavelengths, lower excitation potential, oscillator strength and damping constants (radiative, Stark and Van der Waals) in gfall.dat against more accurate and/or more recent critically evaluated laboratory determinations when available. Specifically, two atomic databases were searched for improved values of these parameters and their uncertainties: the VALD\footnote{http://ams.astro.univie.ac.at/vald/, \cite{1999A&AS..138..119K}} database and the  NIST\footnote{http://physics.nist.gov/cgi-bin/AtData/lines-form} database. We then modified their values in the original linelist accordingly. Damping constants not available in linelists are calculated in SYNSPEC48 using approximations.
We have also excluded lines in the overlaping regions of two successive orders, as we felt that the observed line profiles may not be reliable there. The final linelist contains 270 transitions for 18 elements which we believe are reliable enough to derive the abundances. 
Most of the lines studied here are weak lines which are formed deep in the
atmosphere where LTE should prevail. They are well suited for
abundance determinations.
The final linelist appears in Table \ref{linelist} where, for each element, the wavelength, adopted oscillator strength, its accuracy (when available) and original bibliographical reference are given. We have also included data for hyperfine splitting for the selected transitions when relevant, in particular for Mn II (these were retrieved from the linelist gfhyperall.dat\footnote{http://kurucz.harvard.edu/LINELISTS/GFHYPERALL/}). However the moderate spectral resolution of the spectra and smearing out of spectra by stellar rotation clearly prevent us from detecting signatures of hyperfine splitting and isotopic shifts in our spectra.

\subsection{Synthetic spectra}

Two codes have been used to derive abundances: Takeda's code (\cite{1995PASJ...47..287T}) and SYNSPEC (\cite{1992A&A...262..501H}) to check the abundances produced by Takeda's iterative procedure. We briefly review the assumptions of these codes in section \ref{takeda} and section \ref{synspec}.

\subsubsection{Takeda's procedure}
\label{takeda}

\begin{figure}
\vskip0.7cm
\centering
\includegraphics[width=7cm,height=5.5cm]{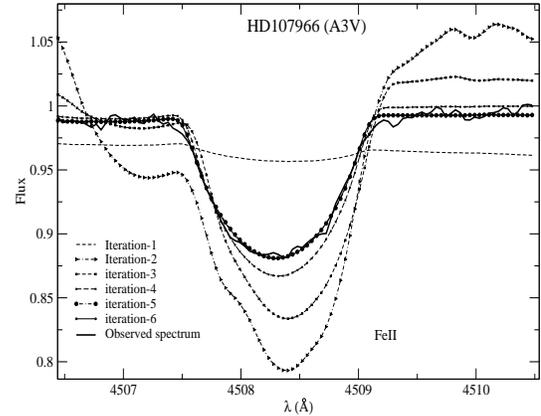}
\caption{Iterative adjustment of synthetic profiles of the FeII line at 4508.288 \AA \ for the A3V star HD107966 with Takeda's program. Convergence was achieved properly at the fifth iteration.}
\label{iteration}
\end{figure}

Takeda's procedure iteratively minimizes the dispersion $\sigma$ between the normalized synthetic spectrum and the observed one defined as: 

\begin{center}
$\sigma^{2}=\sum_{i=1}^{N}\frac{(y_{i}-\eta_{i}-C)^{2}}{N}$
\end{center}  

where $N$ is the number of wavelength points, $y_{i}$ the logarithmic of the observed spectra ($f_{\lambda_{i}}$) and $\eta_{i}$ the logarithmic of the synthetic spectra ($F_{\lambda_{i}}$). C is an offset constant reflecting a possible difference of units between $F_{\lambda}$ and $f_{\lambda}$ (it should be very close to zero when working with normalized fluxes). The synthetic flux $\eta_{i}$ in a line is a priori a function of several physical variables $(x_{1},.....x_{K})$ which represent the unknowns one might be looking for: abundances of individual elements, projected rotational velocity and microturbulent velocity, oscillator strengths, damping constants. The dispersion $\sigma^{2}$ is thus a function of K+1 variables (including C). Minimizing $\sigma^{2}$ thus requires that its partial derivatives 
with respect to these K+1 variables be zero, which leads to a set of K+1 non-linear equations with K+1 unknowns. Takeda (1995) solves this problem numerically using a Newton-Raphson iterative technique (linearization method). Iterations are repeated until convergence ie. when variations in the $x_{k}$s become sufficiently small, typically less than $10^{-3}$.
\\
Takeda's code consists of two routines: the first is a modified version of Kurucz's Width9 code (\cite{1992RMxAA..23...45K}) for computing the opacity data which  needs a Kurucz's model atmosphere as input. Once the opacities are computed, the second routine computes the emergent flux and minimizes the dispersion between the synthetic and the observed spectra. 
\\
In this study, we kept the number of free variables at 3. At each iteration, the code outputs 3 parameters: the abundance(s) of the studied element(s) ($\log \epsilon$), the microturbulent velocity ($\xi_{t}$) constant whith depth and the projected rotational velocity ($v_{e}\sin i$), the oscillator strengths and damping constants being kept fixed. We usually synthesized  unblended lines of one chemical element only, although the procedure allows us to simultaneously synthesize lines of different elements.\\ 
Before tackling the abundance of each individual chemical element, we derived the rotational and microturbulent velocities for each star as follows.
Allowing small variations around solar abundances of Mg and Fe, we iteratively fitted the unblended line profile of the MgII triplet at 4480 \AA \ and a set of neighboring unblended weak and moderately strong FeII lines such as 4491.405 \AA \ and 4508.288 \AA \ leaving $\xi_{t}$ and $v_{e}\sin i$ as free parameters. Even in the fastest rotators, the continuum is well seen in this spectral region where there are only a few regularly spaced lines. The weakest FeII lines are sensitive to rotational velocity and not to microturbulent velocity, the moderately strong FeII lines respond to microturbulent velocity
changes. The MgII triplet, commonly used to derive stellar rotational velocities, appears to be sensitive to both $\xi_{t}$ and $v_{e}\sin i$. Each line of Fe and Mg yielded a set of values for $\log \epsilon$, $\xi_{t}$ and $v_{e}\sin i$, which were in good agreement. For the F stars, the found $\xi_{t}$  were checked against Nissen's (1981) prediction which invoked a polynomial fit for $\xi_{t}$ as a function of $T_{\rm{eff}}$ and $\log g$ for the F stars.
The agreement is generally good, the maximum difference we found being $\Delta\xi_{t}=0.32$ km$\cdot$s$^{-1}$. Morever, the derived 
$v_{e}\sin i$ in this analysis correlate well with those derived from AURELIE spectra of the region around 4500 \AA\ published in \cite{2004IAUS..224..209M}.
\\
For the slowest rotating Am stars, curves of growths of the Fe II lines were also performed. They led to values of $\xi_{t}$ consistent with
those derived from the fit of the Mg II triplet and Fe II lines around 4500 \AA.
Once $\xi_{t}$ and $v_{e}\sin i$ were fixed, we then proceeded to determine the abundance for each selected unblended line of each chemical element. In practice, complete convergence was reached for each star after up to 10
iterations (five in the most favorable cases).
\\ 
A final test on Procyon was also performed, assuming a solar composition (\cite{1985A&AS...59..403S}). The iterative fitting of the Mg II and Fe II lines around 4500 \AA\ yielded an  apparent rotational velocity of 6 km$\cdot$s$^{-1}$ and a microturbulent velocity of 2.2 km$\cdot$s$^{-1}$ in good agreement with Steffen's (\cite{1985A&AS...59..403S}) values ($v_{e}\sin i$=4.5 km$\cdot$s$^{-1}$ and $\xi_{t}$=2.1 km$\cdot$s$^{-1}$ respectively). \\
 Figure \ref{iteration} shows an example of the iterative fitting of the FeII line at 4508,288 \AA\ for the A3V star HD 107966 (6 iterations). For this star, the convergence was already achieved at the fifth iteration. 
 

\subsubsection{Validation of the abundances by SYNSPEC}
\label{synspec}

\begin{figure}
   \centering
   \vskip0.9cm
  \includegraphics[width=7cm,height=6cm]{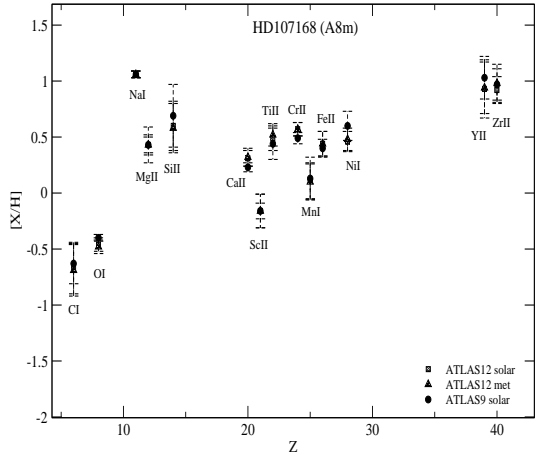}
      \caption{Influence of the underlying model atmosphere on the derived abundances for HD107168 (A8m): black dots are abundances derived using a solar ATLAS9 model, hatched black squares using a solar ATLAS12 model and black triangles are abundances derived with a ATLAS12 model computed for the specific chemical composition of this star.}
         \label{atlas9-12-12m}
	 \end{figure}

We have also used Version 48 of SYNSPEC (\cite{1992A&A...262..501H}) to check the abundances produced by Takeda's iterative procedure. SYNSPEC48 allows us to calculate line profiles of elements up to Z=99. In its LTE mode, SYNSPEC needs a model atmosphere and a linelist plus an auxiliary file containing non standard flags. For a given star, the same ATLAS9 models, the same linelist and the $v_{e}\sin i$ and $\xi_{t}$ derived using Takeda's (1995) procedure were used to calculate synthetic spectra. Only the abundance of the studied unblended line were left as free parameters. The derived abundances with SYNSPEC were found to always agree with those derived from Takeda's procedure within the error bars. The abundances listed in Tables
\ref{abondances-A} and \ref{abondances-F} are those derived from Takeda's procedure.
\\
We also checked the influence of the underlying atmospheric structure on the derived abundances using ATLAS12 (\cite{2005MSAIS...8...14K}). Abundances can be adjusted individually in ATLAS12 which employs the Opacity Sampling technique for line opacity. The effect should be noticeable for the Am stars whose abundances depart most from the solar values.
For Am stars we found that the inclusion of an ALTAS12 model calculated for the specific chemical composition found with ATLAS9 and Takeda's procedure yields new mean abundances which differ by up to 0.08 dex from those derived with ATLAS9. This is less than the error bar on mean abundances. Figure \ref{atlas9-12-12m} compares the abundances of several elements for the A8m star HD107168 derived using ATLAS9 and ATLAS12 model atmospheres. For the other stars, whose chemical compositions depart less from solar, the effect will be even smaller. Since we intend to model all stars in a uniform manner and because of the much larger computing time
needed to run an ATLAS12 model atmosphere, we decided to use only ATLAS9 model atmospheres.  

\section{Results and uncertainties}

\label{resultats}
Abundances for the 22 stars are collected in Table \ref{abondances-A} for A stars and in Table \ref{abondances-F} for F stars. These abundances are relative to the sun ($[\frac{X}{H}]=\log(\frac{X}{H})_{\star}-\log(\frac{X}{H})_{\odot}$)\footnote{Solar abundances are those from \cite{1998SSRv...85..161G}.}. They are weighted means of the abundances derived from each transition. The calculation of their uncertainties and the weighted mean abundances is explained in Appendix A.
\\

\begin{table*}
\caption{Abundances relative to hydrogen and to the solar value, $[\frac{X}{H}]=\log(\frac{X}{H})_{\star}-\log(\frac{X}{H})_{\odot}$ for the A stars. The solar values are those of \cite{1998SSRv...85..161G}. The HD numbers in italics are those for which the uncertainties have been calculated as explained in Appendix A. For the others, the quantities labeled as $\sigma$ are standard deviations.}
\label{abondances-A}
\centering 
 \begin{tabular}{cccccccccccc} 
\hline
HD   	   & SpT 	    &CI	  &$\sigma_{C}$	   &OI   &$\sigma_{O}$	 & NaI   &$\sigma_{Na}$   &MgII   &$\sigma_{Mg}$    &SiII    
&$\sigma_{Si}$ \\ \hline \\
\it{HD107966}   &A3V/A3IV	& -0.07   & 0.04  & 0.00  &  0.05   & 0.52  &  0.13  &0.13  &  0.09 & 0.14   &   0.09		    \\
HD108382   	&A4V/A3IV	& -0.71   & 0.15  & -0.02 &  0.06   & 0.24  &  0.18  &0.13  &  0.12 & 0.18   &   0.19		    \\
HD106887   	&A4m 	 	& -0.56   & 0.15  & -0.43 &  0.06   & 0.54  &  0.08  &0.17  &  0.13 & 0.23   &   0.20		   \\
\it{HD107655}   &A0V 	 	& -0.75   & 0.19  & -0.58 &  0.06   & 0.55  &  0.21  &-0.12 &  0.18 & 0.06   &   0.07		    \\
\it{HD107168}   &A8m/kA5hA5mF0	& -0.65   & 0.08  & -0.40 &  0.04   & 1.06  &  0.03  &0.49  &  0.18 & 0.77   &   0.12		   \\
HD109307   	&A4Vm/A3IV-V  	& -0.36   & 0.01  & -0.31 &  0.12   & 0.65  &  0.45  &-0.07 &  0.29 & 0.23   &   0.18		   \\
HD108642   	&A2m/kA2hA7mA7	& -0.75   & 0.16  & -0.85 &  0.12   & 0.13  &  0.14  &0.25  &  0.19 & 0.09   &   0.09		   \\
HD107276   	&Am/Ka5mA7	& -0.13   & 0.27  & -0.08 &  0.04   & 0.51  &  0.08  &0.25  &  0.18 & 0.21   &   0.18		    \\
HD108486   	&AmkA3hA5mA7  	& -0.69   & 0.20  & -0.86 &  0.04   & 0.60  &  0.34  &-0.02 &  0.18 & 0.02   &   0.21		   \\  
HD106999   	&Am  	 	& -0.15   & 0.10  & -0.02 &  0.04   & 1.04  &  0.11  &0.40  &  0.18 & 0.26   &   0.09		   \\
\it{HD107513}   &Am/kA7hF0mF0 	& -0.36   & 0.10  & -0.15 &  0.12   & 0.10  &  0.10  &0.22  &  0.16 & 0.12   &   0.12		   \\

\hline \hline
HD   	   & SpT 	    &CaII        &	$\sigma_{Ca}$   &ScII 	   &$\sigma_{Sc}$	&TiII	  &$\sigma_{Ti}$ &VII   &$\sigma_{V}$ &CrII   
&$\sigma_{Cr}$ \\ \hline \\
\it{HD107966}   &A3V/A3IV	& -0.08 &  0.08 &  -0.13  &  0.08   & -0.10  & 0.05   &-      & -      &   0.08	& 0.09  \\
HD108382   	&A4V/A3IV	& -0.19 &  0.04 &  0.12   &  0.26   & -0.09  & 0.17   & -     &-       &   0.24	& 0.25 \\
HD106887   	&A4m 	 	& -0.11 &  0.21 &  0.03   &  0.34   & 0.12   & 0.15   & -     &-       &   0.13	& 0.12 \\
\it{HD107655}   &A0V 	 	&- 	&-	&  -0.20  &  0.11   & 0.02   & 0.08   &-      &-       &   0.56	& 0.09 \\
\it{HD107168}   &A8m/kA5hA5mF0	& 0.23  &  0.16 &  -0.18  &  0.10   & 0.48   & 0.08   &-      & -      &   0.48	& 0.20 \\
HD109307   	&A4Vm/A3IV-V  	& 0.14  &  0.06 &  0.10   &  0.11   & -0.07  & 0.10   &-      &-       &   0.03	& 0.05 \\
HD108642   	&A2m/kA2hA7mA7	& -0.33 &  0.18 &  -1.17  &  0.20   & -0.18  & 0.07   &0.86  &  0.13 &   0.18	& 0.05 \\
HD107276   	&Am/Ka5mA7	& -0.24 &  0.16 &  -0.06  &  0.23   & 0.04   & 0.22   &0.90  &  0.10 &   -0.08  & 0.11  \\
HD108486   	&AmkA3hA5mA7  	& -0.22 &  0.15 &  -0.50  &  0.31   & -0.20  & 0.13   &0.70  &  0.29 &   0.08	& 0.15  \\  
HD106999   	&Am  	 	& 0.10  &  0.16 &  0.41   &  0.27   & 0.14   & 0.21   &0.69  &  0.29 &   0.03	& 0.20  \\
\it{HD107513}   &Am/kA7hF0mF0 	& -	&-	&  -0.25  &  0.10   & 0.01   & 0.07   &0.65  &  0.32 &   0.09	& 0.09  \\

\hline \hline
HD   	   	& SpT 	    	&MnI  &$\sigma_{Mn}$  &FeII	 &  $\sigma_{Fe}$   &Co	&$\sigma_{Co}$       &NiI      &$\sigma_{Ni}$ &SrII   &$\sigma_{Sr}$ \\ \hline \\
\it{HD107966}   &A3V/A3IV	& -0.35 &   0.18    &-0.13 &   0.05 &  0.65  &   0.38	   &-0.18 &   0.07  &	-0.28&    0.33       \\
HD108382   	&A4V/A3IV	& -0.55 &   0.12    &-0.14 &   0.10 &-	     & -   	   &-0.14 &   0.16  &	-0.26&    0.02      \\
HD106887   	&A4m 	     	& 0.02  &   0.21    &0.21  &   0.15 &-	     & -   	   &0.37  &   0.11  &	0.59 &    0.08      \\
\it{HD107655}   &A0V 	     	& -	& -	    &0.08  &   0.05 &-	     & -   	   &0.77  &   0.23  &	-0.23&    0.20      \\
\it{HD107168}   &A8m/kA5hA5mF0  & 0.19  &   0.12    &0.39  &   0.10 &-	     & -   	   &0.60  &   0.10  &	0.79 &    0.31      \\
HD109307   	&A4Vm/A3IV-V    & -0.04 &   0.17    &0.05  &   0.09 &  -0.06 &   0.35	   &0.12  &   0.08  &	0.46 &    0.10      \\
HD108642   	&A2m/kA2hA7mA7  & -0.04 &   0.10    &0.16  &   0.11 &  0.29  &   0.35	   &0.41  &   0.08  &	0.50 &    0.13      \\
HD107276   	&Am/Ka5mA7	& -	&- 	    &0.03  &   0.23 &-	     &-    	   &-0.14 &   0.10  &	-0.22&    0.19       \\
HD108486   	&AmkA3hA5mA7    & -	&- 	    &0.20  &   0.09 &-	     &-    	   &0.21  &   0.21  &	0.76 &    0.02       \\ 
HD106999   	&Am  	     	& -	&- 	    &0.08  &   0.13 &-	     &-    	   &0.23  &   0.20  &	0.47 &    0.07       \\
\it{HD107513}   &Am/kA7hF0mF0   & -	&- 	    &-0.02 &   0.07 &-	     &-    	   &-0.23 &   0.10  &	-0.06&    0.19       \\

\hline \hline 
HD   	   	& SpT 	    	&YII  &$\sigma_{Y}$  &ZrII	 &  $\sigma_{Zr}$   &BaII	&$\sigma_{Ba}$         & \\ \hline \\
\it{HD107966}   &A3V/A3IV	&0.00	& 0.13 &   0.38  &  0.12     &0.04  &  0.26  	  \\
HD108382   	&A4V/A3IV	&0.25	& 0.21 &   0.16  &  0.17     &-0.24 &  0.03  	 \\
HD106887   	&A4m 	     	&0.78	& 0.16 &   0.61  &  0.15     &1.40  &  0.26  	 \\
\it{HD107655}   &A0V 	     	&0.80	& 0.15 &   0.78  &  0.12     &0.72  &  0.32  	 \\
\it{HD107168}   &A8m/kA5hA5mF0  &0.97	& 0.12 &   0.93  &  0.13     &      &	     	 \\
HD109307   	&A4Vm/A3IV-V    &0.56	& 0.07 &   0.60  &  0.11     &1.17  &  0.23  	 \\
HD108642   	&A2m/kA2hA7mA7  &0.88	& 0.12 &   0.75  &  0.12     &1.79  &  0.24  	 \\
HD107276   	&Am/Ka5mA7	&0.12	& 0.10 &   -0.15 &  0.20     &0.44  &  0.21  	  \\
HD108486   	&AmkA3hA5mA7    &0.56	& 0.12 &   0.67  &  0.13     &1.57  &  0.21  	  \\ 
HD106999   	&Am  	     	&0.27	& 0.11 &   0.49  &  0.08     &0.78  &  0.07  	  \\
\it{HD107513}   &Am/kA7hF0mF0   &0.08	& 0.11 &   0.13  &  0.20     &0.53  &  0.15  	  \\

\hline \hline \\	 	  	        										
\end{tabular}
 \end{table*}

\begin{table*}
\caption{Abundances relative to hydrogen and to the solar value, $[\frac{X}{H}]=\log(\frac{X}{H})_{\star}-\log(\frac{X}{H})_{\odot}$ for the F stars. The HD numbers in italics are those for which the uncertainties have been calculated as explained in Appendix A. For the others, the quantities labeled as $\sigma$ are standard deviations.}
\label{abondances-F}
\centering 
 \begin{tabular}{cccccccccccc} 
\hline
HD   	   & SpT 	    &CI	  &$\sigma_{C}$	   &OI   &$\sigma_{O}$	 & NaI   &$\sigma_{Na}$   &MgII(MgI)   &$\sigma_{Mg}$    &SiII    
&$\sigma_{Si}$ \\ \hline \\
\it{HD106103}   &F5V 	&-0.04  &    0.08 &  -0.32  &    0.15  & -0.04  &    0.04  &0.30(0.02)    &  	0.21(0.09)&  0.10  &     0.10	\\
HD106293   &F5V    	&-0.07  &    0.27 &-	    &  -        & 0.01   &    0.18  &0.45(0.10)    &  	0.18(0.06)&  0.26  &     0.11\\
HD106691   &F5IV   	&-0.07  &    0.16 &  -0.26  &    0.15  & -0.06  &    0.20  &0.30(0.10)    &  	0.18(0.07)&  0.12  &     0.08\\
HD106946   &F2V    	&-0.02  &    0.20 &  0.02   &    0.15  & 0.09   &    0.14  &0.45(0.13)    &  	0.18(0.04)&  0.26  &     0.07\\
\it{HD107611}   &F6V    &0.05   &    0.10 &-	    & -         & -0.10  &    0.04  &0.23(0.10)    &  	0.18(0.15)&  0.22  &     0.12\\
\it{HD109530}   &F2V    &0.08   &    0.09 &  0.10   &    0.14  & 0.16   &    0.12  &0.39(-0.01)   &  	0.11(0.13)&  0.13  &     0.28\\
HD107877   &F6     	&-      & -  	  &-	    &-          & -0.13  &    0.03  &0.40	  &  	0.18	  &-	   &-  	     \\
HD108154   &F5     	&-       &-   	  &-	    &-          & -0.13  &    0.21  &0.15	  &  	0.18	  &-	   &-  	    	\\
HD108226   &F5     	&-       &-   	  &-	    &-          & 0.04   &    0.23  &0.22	  &  	0.18	  &-	   &-  	    	\\ 
HD108976   &F6 V   	&-       &-   	  &-	    &-          & 0.02   &    0.20  &0.13	  &  	0.18	  &-	   &-  	    	\\
HD109069   & F0 V  	&-       &-   	  &-	    &-          &     -   & -  	   &0.41	  &  	0.18	  &-	   &-  	    	\\

\hline \hline
HD   	   & SpT 	    &CaII        &	$\sigma_{Ca}$   &ScII 	   &$\sigma_{Sc}$	&TiII	  &$\sigma_{Ti}$ &VII   &$\sigma_{V}$ &CrII   
&$\sigma_{Cr}$ \\ \hline \\
\it{HD106103}   &F5V 	&-0.17 &  0.21 & 0.00	& 0.08 &0.00  &  0.10 &0.44 &	0.15&  0.05 &	0.10   \\
HD106293   &F5V    	&-      & -      & -0.07  & 0.15 &0.01  &  0.16 &0.62 &	0.19&  0.11 &	0.09\\
HD106691   &F5IV   	&-0.23 &  0.19 & -0.01  & 0.16 &0.00  &  0.15 &0.42 &	0.33&  0.04 &	0.12\\
HD106946   &F2V    	&-      & -      & -0.04  & 0.17 &0.21  &  0.17 &0.60 &	0.44&  0.11 &	0.18\\
\it{HD107611}   &F6V    &-      &-       & -0.05  & 0.08 &0.14  &  0.06 &0.68 &	0.10&  0.10 &	0.08\\
\it{HD109530}   &F2V    &-      &-       & 0.06	& 0.14 &0.25  &  0.10 &0.45 &	0.33&  0.02 &	0.08\\
HD107877   &F6     	&-      &-       &-	&-      &0.18  &  0.23 &0.64 &	0.26&  0.33 &	0.08\\
HD108154   &F5     	&-      &-       &-	&-      &0.18  &  0.13 &0.12 &	0.36&  0.38 &	0.08   \\
HD108226   &F5     	&-      &-       &-	&-      &0.24  &  0.14 &0.57 &	0.25&  0.38 &	0.08   \\ 
HD108976   &F6 V   	&-      &-       &-	&-      &0.09  &  0.15 &0.56 &	0.12&  0.22 &	0.08   \\
HD109069   & F0 V  	&-      &-       &-	&-      &-0.01 &  0.21 &0.40 &	0.02&-	    &-	       \\

\hline \hline
HD   	   	& SpT 	    	&MnI  &$\sigma_{Mn}$  &FeII	 &  $\sigma_{Fe}$   &Co	&$\sigma_{Co}$       &NiI      &$\sigma_{Ni}$ &SrII   &$\sigma_{Sr}$ \\ \hline \\
\it{HD106103}   &F5V   	&-0.06 &  0.07       &0.09   & 0.05   & -0.01&   0.26 & 0.13  &  0.07& 0.20  &  0.15  \\	 
HD106293   &F5V   	&0.04  &  0.09       &0.19   & 0.17   & 0.44 &	 0.26 & 0.21  &  0.22& -0.01 &  0.00\\
HD106691   &F5IV  	&0.08  &  0.10       &-0.08  & 0.16   & 0.64 &	 0.14 & -0.04 &  0.08& 0.15  &  0.02 \\
HD106946   &F2V   	&0.08  &  0.07       &0.18   & 0.08   & 0.29 &	 0.25 & -0.02 &  0.18& 0.12  &  0.10 \\
\it{HD107611}   &F6V   	&0.01  &  0.06       &0.09   & 0.05   & -0.10&   0.33 & -0.06 &  0.05& 0.14  &  0.17 \\
\it{HD109530}   &F2V   	&-0.20 &  0.31       &0.15   & 0.09   & -0.02&   0.24 & 0.21  &  0.11& 0.27  &  0.20\\
HD107877   &F6    	&0.15  &  0.12       &0.15   & 0.13   & -0.30&   0.28 & 0.12  &  0.08 &-     &-      \\
HD108154   &F5    	&0.13  &  0.13       &0.01   & 0.11   & -0.35&   0.30 & 0.09  &  0.08 &-     &-       \\
HD108226   &F5    	&0.11  &  0.07       &0.05   & 0.11   & -0.30&   0.34 & 0.08  &  0.08 &-     &-       \\
HD108976   &F6 V  	&-0.46 &  0.12       &-0.01  & 0.14   & -0.29&   0.16 & 0.03  &  0.13 &-     &-       \\
HD109069   & F0 V 	&-      &- 	     &-0.03  & 0.17   & 0.45 &	 0.37 & 0.29  &  0.14 &-     &-       \\

\hline \hline
HD   	   & SpT 	&YII	&$\sigma_{Y}$ &ZrII&$\sigma_{ZrII}$&BaII&$\sigma_{Ba}$ 	 \\ \hline \\
\it{HD106103}   &F5V   	&0.00	& 0.08   &0.04   & 0.11 & 0.86  &  0.10    \\	    
HD106293   &F5V   	&0.01	& 0.06   &-0.04  & 0.09 & 0.64  &  0.23    \\
HD106691   &F5IV  	&0.04	& 0.08   &0.50   & 0.15 & 0.65  &  0.21    \\
HD106946   &F2V   	&0.24	& 0.19   &0.38   & 0.09 & 0.59  &  0.16    \\
\it{HD107611}   &F6V   	&0.10	& 0.12   &0.13   & 0.17 & 0.84  &  0.10   \\
\it{HD109530}   &F2V   	&0.06	& 0.20   &-0.13  & 0.14 & 0.43  &  0.14  \\
HD107877   &F6    	&-	&-	 &0.14   & 0.17 & 0.56  &  0.10   \\
HD108154   &F5    	&-	&-	 &0.03   & 0.17 & 0.54  &  0.10    \\
HD108226   &F5    	&-	&-	 &-0.06  & 0.17 & 0.59  &  0.10    \\
HD108976   &F6 V  	&-	&-	 &-0.50  & 0.17 & 0.29  &  0.10    \\
HD109069   & F0 V 	&-	&-	 &-0.04  & 0.17 & 0.37  &  0.10    \\

\hline \hline

\end{tabular}
 \end{table*}


\subsection{The general abundance pattern of A, Am and F stars in Coma Ber}

\begin{figure*}
\vskip0.7cm
\centering
\begin{tabular}{cc}
\includegraphics[scale=0.4]{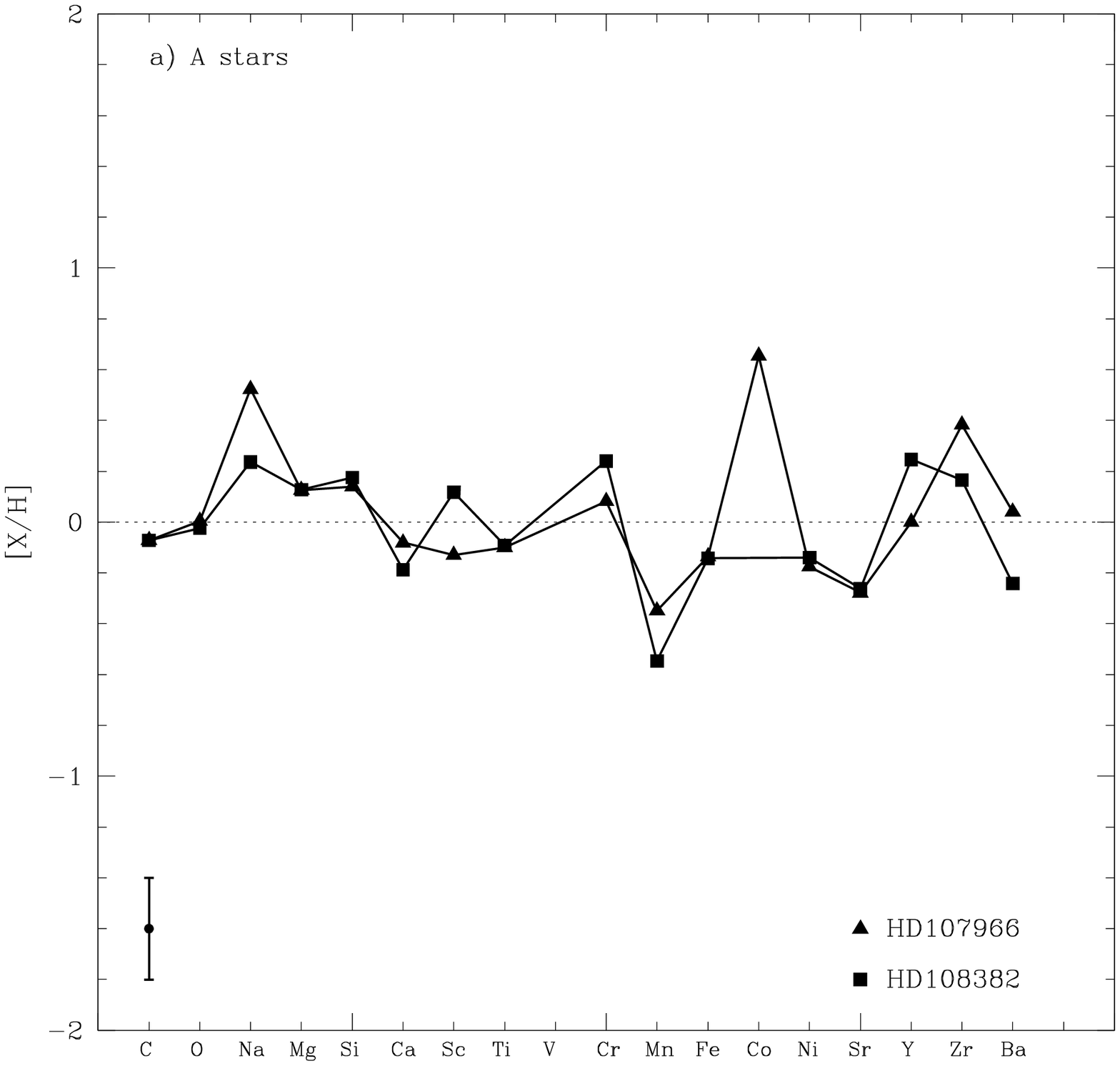}&
\includegraphics[scale=0.4]{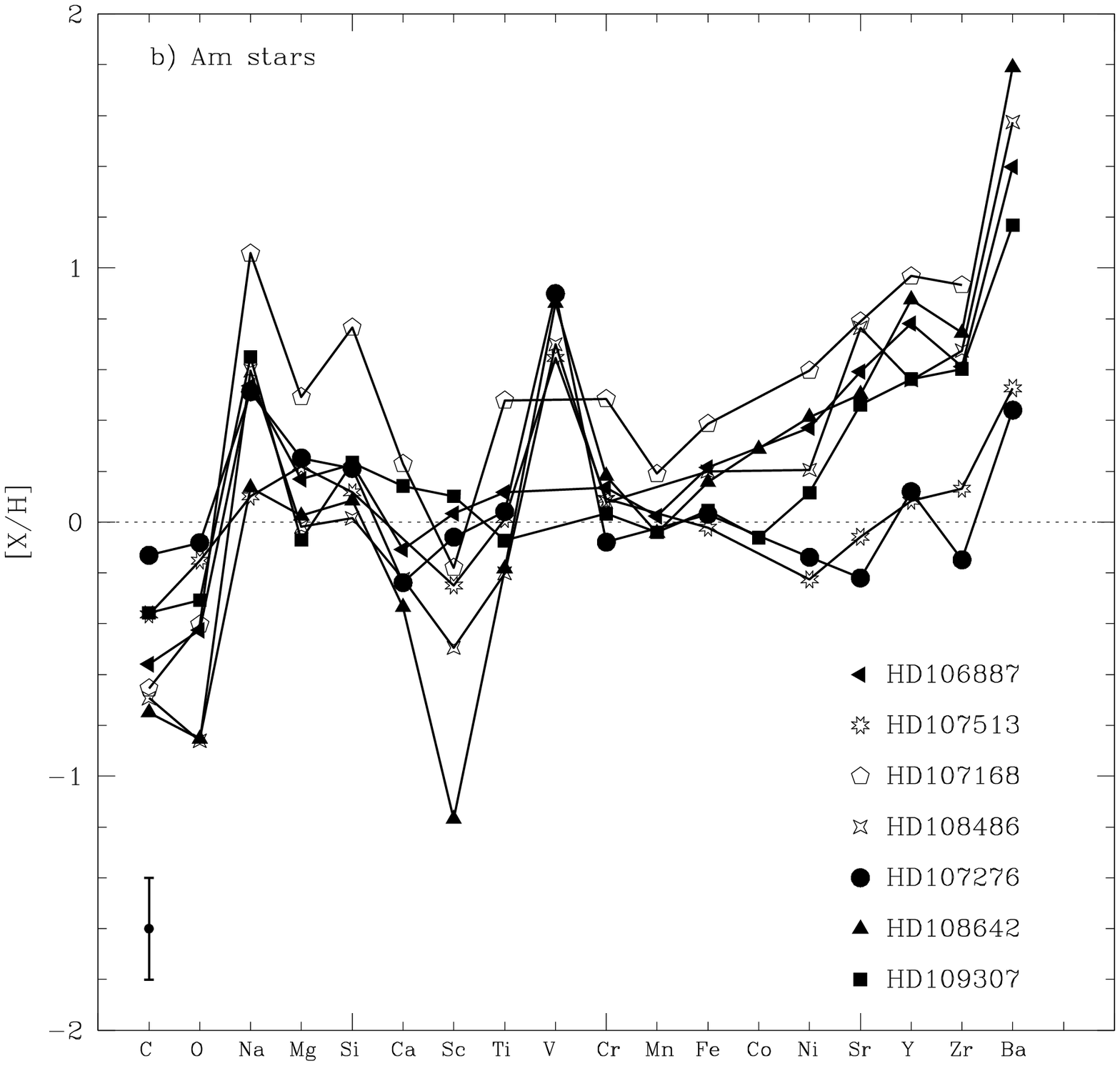}\\ \\ \\ \\  
\includegraphics[scale=0.4]{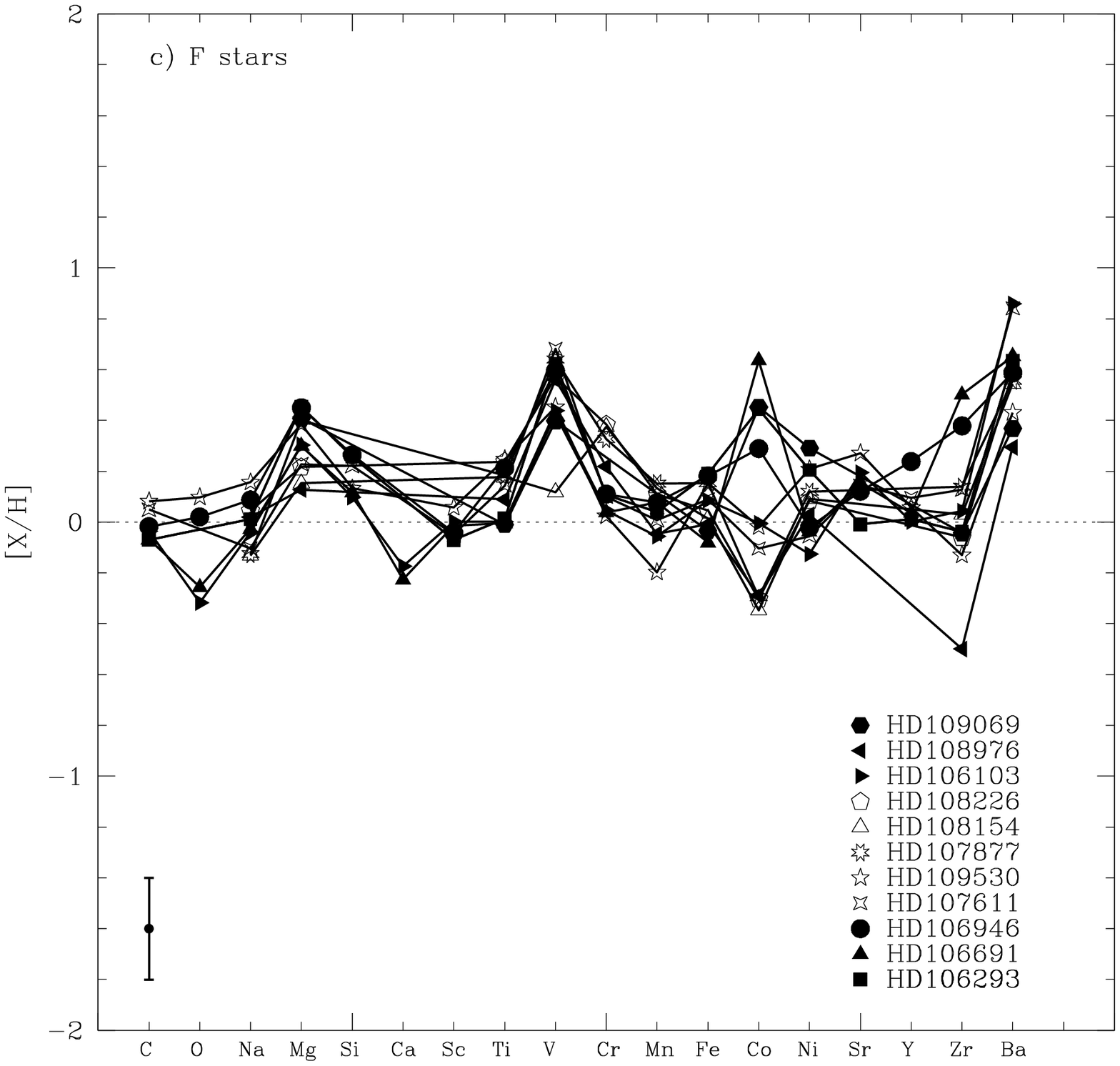}
\end{tabular}
\caption{Abundance patterns for the "normal" A (a), Am (b) and F stars of Coma Berenices cluster. A maximum $\pm$0.30 dex error bar is displayed. The horizontal dashed line represent the solar composition.}
\label{abond-A-Am-F}
 \end{figure*}

Graphs where abundances are displayed against atomic number (abundance pattern plots) are particularly appropriate to compare the behaviour of the A, Am and F stars for different chemical elements. The abundance patterns for the A stars, Am stars and F stars are displayed respectively in Figures \ref{abond-A-Am-F}a, \ref{abond-A-Am-F}b and \ref{abond-A-Am-F}c.
\\
The seven Am stars of the clusters display a characteristic jig-saw pattern with much larger excursions from the solar compositions than the normal A stars do. This trend had already been found by Hui-Bon-Hoa \& Alecian (1998) who derived the abundances of Mg, Ca, Sc, Cr, Fe and Ni in 
four Am stars of our sample (HD107168, HD107966, HD108486 and HD109307). 
Our abundances for these elements agree well with theirs for HD 107168, for which all fundamental parameters are similar in both studies. However, slight to moderately large differences occur for HD108486, HD109307 and HD107966, for which we adopted larger microturbulent velocities (in particular for HD107966), all other parameters being consistent in both studies.  
The 2 normal A stars of our sample, HD107966 and HD108382, have very similar abundances, nearly solar, in almost all elements and thus exhibit almost the same abundance pattern (Figure \ref{abond-A-Am-F}a).
Inspection of Figure \ref{abond-A-Am-F}b reveals that generally, for almost all chemical elements, Am stars display star-to-star variations that can be larger than the typical uncertainty. All Am stars are deficient in C and O, but not all are deficient in Ca and/or Sc and most have pronounced overabundances of iron-peak and heavy elements. The two normal A stars have almost solar abundances in most elements. The F stars definitely exhibit less scatter than the Am stars. They tend to be mildly overabundant, in particular in Mg, Si, V and Ba. They have nearly solar abundances for C, O, Na, Ti, Fe, Ni, Sr and Y.  

\subsection{Behaviour of individual elements}
\label{discussion}

In this section, we use the found abundances for each element to address two issues. 
First, do the abundances depend on stellar parameters such as the effective temperature and $v_{e}\sin i$ ? Any such correlation could be very valuable for theorists investigating the various hydrodynamical mechanisms affecting photospheric abundances. Second, 
how do the abundances of each element vary with respect to each other? Given that iron is astrophysically one of the most important elements (since it provides a rough estimate of the "metallicity") and given that the abundances derived for this element are probably the most reliable, we have examined whether the abundances of individual elements correlate with those of iron. Hill (1995) and Lemke (1989, 1990) also looked for similar correlations and we will refer to their findings later.
\\
We can roughly separate the elements we studied into two groups. For most elements (Fe II , Ti II, OI, Cr II, Mg II, Mn I, C I, Si II, Ca II and Ni I), we synthesized several lines of quality A to D and we can be confident in the abundances we derived, in particular for iron, titanium and chromium. For elements with many lines, errors in individual oscillator strengths should tend to cancel out. For other elements, we have very few lines (Sr II), so that errors on oscillator strengths may induce scatter in the abundances. For
V II, Co I, Y II and Zr II, several lines are available but their accuracy is not necessarily well known. The abundances of these elements should be viewed with more caution
However, our main goal here is to investigate how the abundances of individual elements vary with that of iron and also to study star-to-star variations of fundamental parameters. This can be established independently of errors in the absolute values of the oscillator strengths, since all stars will be affected in the same manner.

\subsubsection{The light elements} 

\subsubsection*{Behaviour of carbon:}

\begin{figure*}
\centering
\vskip0.5cm
\includegraphics[scale=0.8]{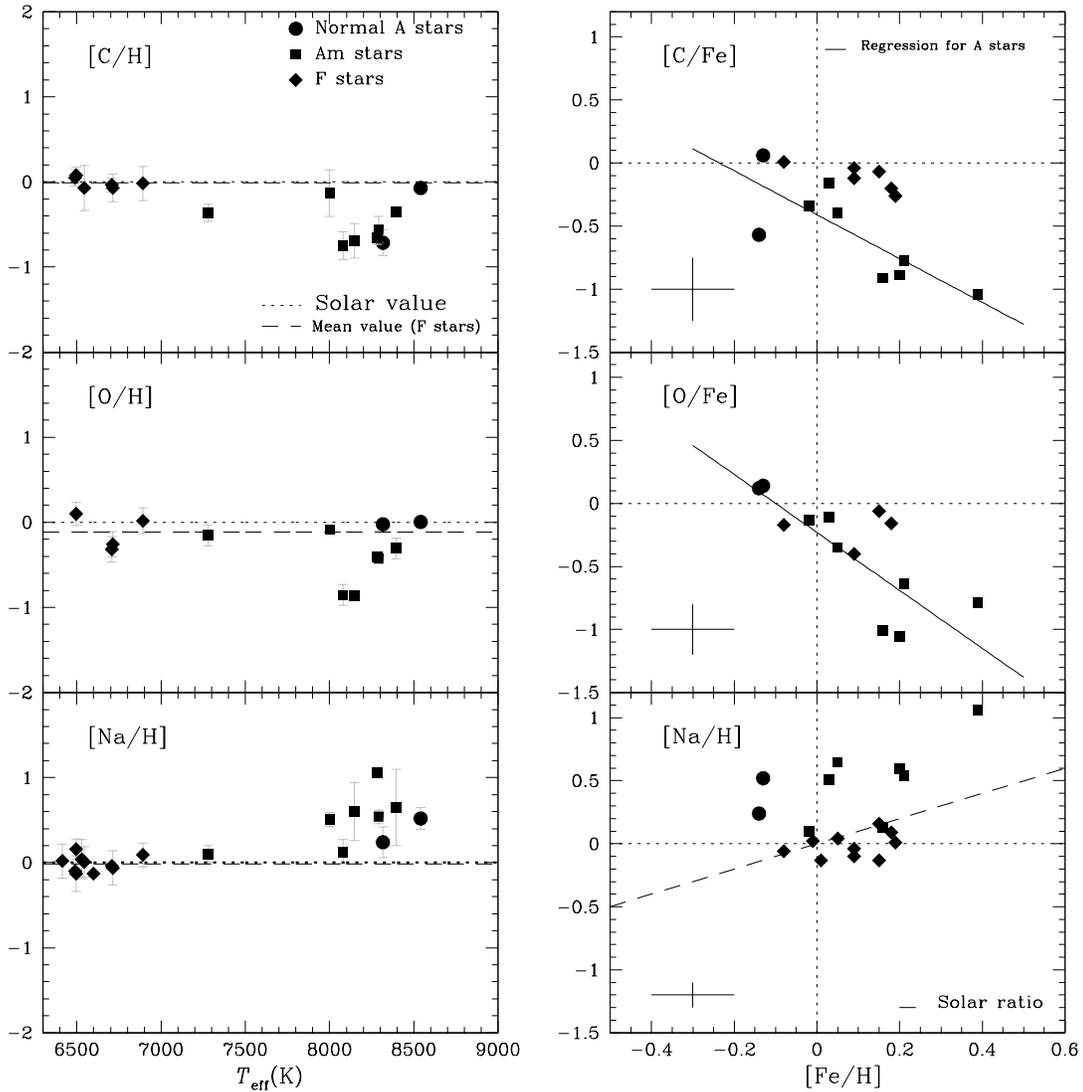}
\caption{Left panel: Abundance of carbon, oxygen and sodium versus effective temperature. The dotted line corresponds to the solar value and the dashed one to the mean value determined for the F stars of the cluster. Right panel: [C/Fe], [O/Fe] and [Na/H] versus [Fe/H]. The filled dots correspond to normal A stars, the filled squares correspond to Am stars and the filled diamonds correspond to F stars. In the plot representing [Na/H] versus [Fe/H], the dashed line corresponds to the solar [Na/Fe] ratio. The error bars in the right panel represent the mean uncertainties for the displayed abundances.}
\label{CCA}
\end{figure*}

Seven lines of quality B of C I have been synthesized. 
In a graph of $[\frac{C}{Fe}]$ versus $T_{\rm{eff}}$, this element displays a different behaviour in F stars and A stars. The F stars show very little scatter around their mean iron abundance (about -0.01 dex) whereas the A and Am stars exhibit a pronounced spread in abundances of about 0.75 dex which is much larger than the estimated typical uncertainty of about 0.15 dex on $[\frac{C}{H}]$. Also all A stars display deficiencies in carbon. There seems thus to be real star-to-star variation in $[\frac{C}{Fe}]$. Another way to quantify the relative dispersion of A and Am stars with respect to the F stars is to compute the mean abundance for carbon and the associated dispersion for each group of stars, $\sigma_{A}$ and $\sigma_{F}$.
For A and F stars, these mean abundances and dispersions for carbon and all other chemical elements are collected in Table \ref{abon-moyenne-deviation}. For carbon, the dispersion $\sigma_{\rm{A}}$ for the A stars is about 4 times higher than for F stars.
\\
The behaviour of carbon with respect to iron is not clear when $[\frac{C}{H}]$ is displayed versus 
$[\frac{Fe}{H}]$. As previously found by Hill (1995),
an anticorrelation appears more clearly when $[\frac{C}{Fe}]$ is displayed against $[\frac{Fe}{H}]$ for Am and normal A stars (Figure \ref{CCA}). We find a slope of $-1.74 \pm 0.48$ , not very different from the value obtained by Hill (1995) ($-1.24\pm0.31$ for 15 A stars).\\
We have checked whether the C lines studied here might be affected by non-LTE effects especially for A stars. Non-LTE abundance corrections of carbon have been calculated by \cite{1996A&A...312..966R} for a set of main sequence stars ranging in effective temperatures from 7000 K to 12000 K, surface gravities from $\log g=3.5$ to 4.5 dex and metallicities from $[\frac{M}{H}]=-0.5$ to 1 dex. For effective temperatures below 10000 K, the non-LTE abundance corrections were found to be always negative. Only three lines of our list were studied by \cite{1996A&A...312..966R}: $\lambda$5052.17, $\lambda$5380.34 and $\lambda$6587.61 \AA. According to her Figures 7 and 8, the corrections for these lines, whose equivalent width are always less than 100 m\AA, should be in the range -0.10 to -0.25 dex for the temperatures and metallicities of the stars in our sample, making these stars even more deficient in carbon. These corrections do not appear to diminish the star-to-star scatter in [C/H].

\subsubsection*{Oxygen}

The derived oxygen abundances are the weighted means of 16 quality B and 6 quality C+ lines.
In a graph of $[\frac{O}{H}]$ versus $T_{\rm{eff}}$ (Figure \ref{CCA}), the F stars abundances are slightly scattered around their mean value. All A and Am stars exhibit underabundances of oxygen and their scatter is again larger than that of the F stars. The total spread in oxygen abundance for the A and Am stars is about 0.8 dex, significantly larger than the maximum uncertainty of 0.12 dex for A stars. This again suggests real star-to-star variations in [O/H].
\\
In a graph of $[\frac{O}{H}]$ versus $[\frac{Fe}{H}]$, oxygen seems to be only very loosely anticorrelated with iron. The anticorrelation appears more clearly in $[\frac{O}{Fe}]$ versus $[\frac{Fe}{H}]$ with a slope of $-2.30 \pm 0.53$. 
\cite{2000A&A...359.1085P} calculated the line formation of most of the O I lines analysed here in A star atmospheres using the codes DETAIL and SURFACE (Giddings 1981). For stars with effective temperatures less than 10000 K and $\log g$ around 4.0, these corrections always remain less than -0.03 dex, well below the estimated uncertainty. The observed anticorrelation of [O/Fe] versus [Fe/H] should therefore remain after non-LTE corrections have been made. The large star-to-star variations in [O/H] would not be affected by these corrections either.

\begin{figure*}
\centering
\vskip0.5cm
\includegraphics[scale=0.8]{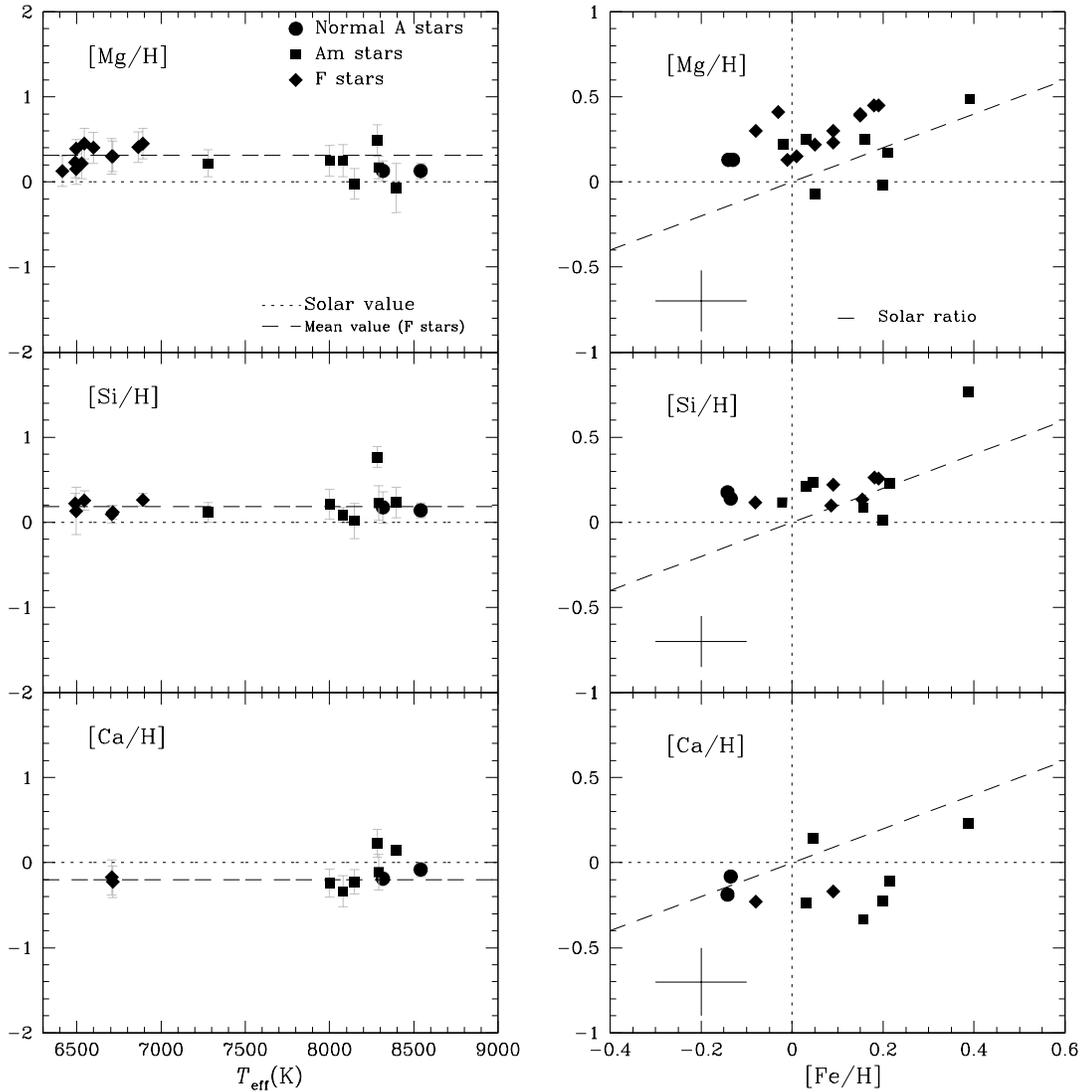}
\caption{Left panel: Abundance of magnesium, silicon and calcium versus effective temperature. The dotted line represents the solar value and the dashed one represents the mean abundance of F stars. Right panel: [Mg/H], [Si/Fe] and [Ca/H] versus [Fe/H]. The symbols are the same as in Figure \ref{CCA}. The dashed lines represent the solar ratios.}
\label{MgSiCa}

\end{figure*}

\subsubsection*{Sodium}
\label{sodium}
Sodium abundances were derived from 9 lines of quality A to C. The sodium abundances for the F stars display very little scatter around their mean value (-0.01 dex), very close to solar. All A and Am stars display overabundances of Na. For these stars, the spread in $[\frac{Na}{H}]$ is about 0.9 dex, much larger than the typical uncertainty (0.10 dex). This suggests there are real star-to-star variations in $[\frac{Na}{H}]$. The sodium abundance does not appear to be correlated to the iron abundance: in Figure \ref{CCA}, about half of the data lie above the line of solar ratio while the other half are scattered around it.
\\
 Quantitative information on non-LTE corrections for Na abundances in A stars is scarce. Non-LTE corrections have been calculated by \cite{2002A&A...389..537B} for two cool A8IV-V stars whose fundamental parameters are close to the coolest A star of our sample. The largest corrections amount to -0.35 dex and occur for the two lines $\lambda$5889.95 \AA\ and 5895.92 \AA. In the absence of calculations for hotter stars with various $[\frac{Fe}{H}]$, it is difficult to predict what the non-LTE corrections would be for the A and Am stars studied here. Provided the corrections would be small for F stars, lowering the A and Am star abundances by about 0.35 dex would certainly improve the correlation of $[\frac{Na}{H}]$ with $[\frac{Fe}{H}]$.

\begin{figure*}
\centering
\vskip0.5cm
\includegraphics[scale=0.8]{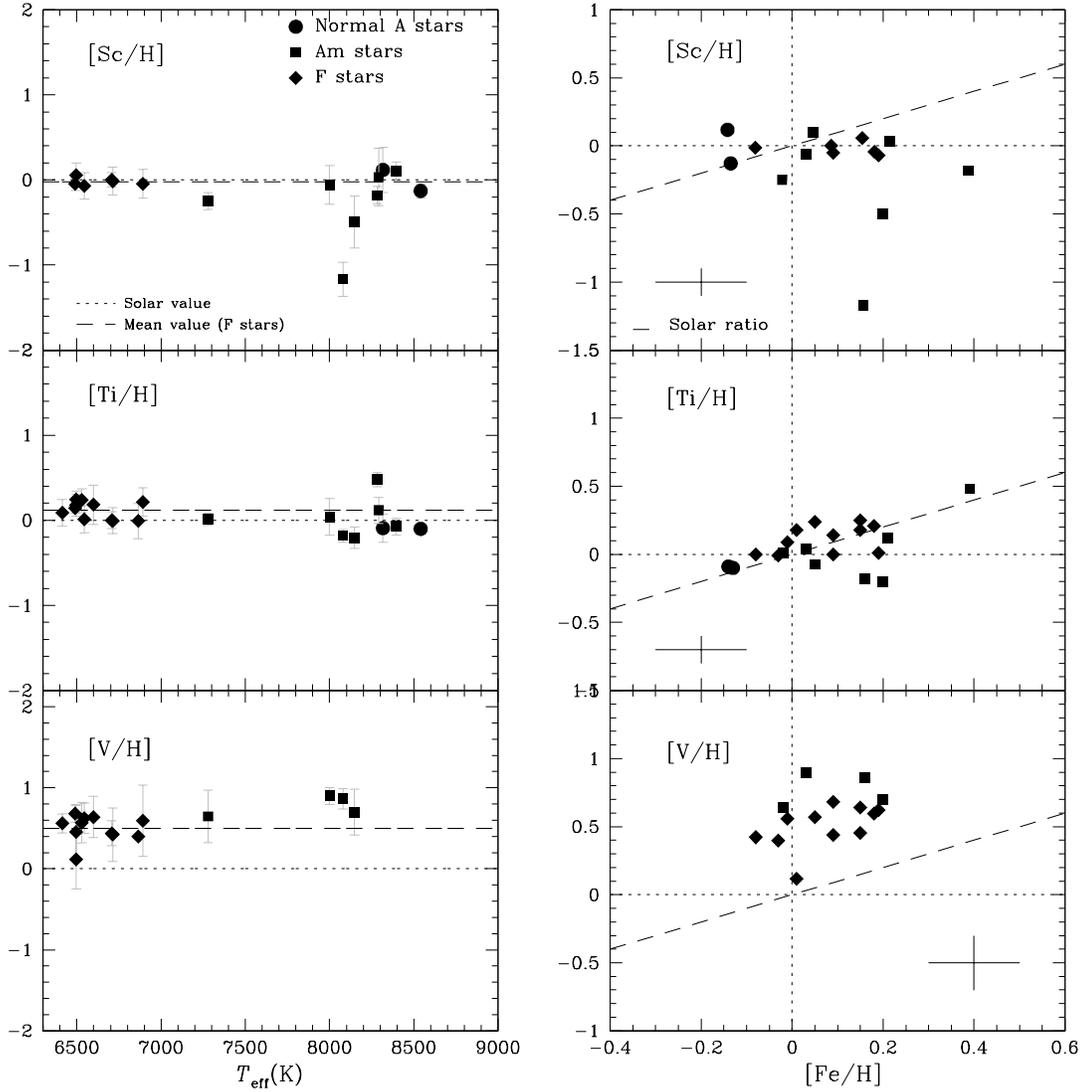}
\caption{Left panel: Abundance of scandium, titanium and vanadium versus effective temperature. The dotted line represents the solar value and the dashed one represents the mean abundance of F stars. Right panel: [Sc/H], [Ti/Fe] and [V/H] versus [Fe/H]. The symbols are the same as in Figure \ref{CCA}. The dashed lines represent the solar ratios.}
\label{ScTiV}
\end{figure*}

\subsubsection*{Magnesium} 

Seven lines of Mg II of quality B to D were used for the A stars and 5 lines of Mg I of qualities B to C for the F stars.
In graphs of $[\frac{Mg}{H}]$ versus $T_{\rm{eff}}$ (Figure \ref{MgSiCa}), the F stars and the A and Am stars display fairly large and comparable scatter. The maximum spread in $[\frac{Mg}{H}]$ is about 0.30 dex and 0.55 dex for the F and the A and Am stars respectively. The maximum uncertainty in $[\frac{Mg}{H}]$ being about 0.18 dex, there does not appear to be significant star-to-star variations. Almost all stars exhibit LTE overabundances. We noticed that the Mg II $\lambda$ 4481 \AA\ triplet systematicaly yields higher abundances than other Mg II lines. Excluding this triplet reduces the abundances by about -0.30 dex for the A and F stars.
\\
The $[\frac{Mg}{H}]$ abundances (derived from all lines excluding the Mg II triplet) do not appear to be correlated to $[\frac{Fe}{H}]$. Most data lie above the line of the solar ratio $[\frac{Mg}{Fe}]$.
In their analysis, Hill \& Landstreet (1993) found that the ratio $[\frac{Mg}{Fe}]$ runs parallel to and above the line of solar $[\frac{Mg}{Fe}]$. They attribute this to the use of the very strong Mg II $\lambda$ 4481 \AA\ triplet (the only line they analysed).\\
\cite{2001A&A...369.1009P} computed a model atom for non-LTE line formation for neutral and singly-ionized magnesium to evaluate non-LTE corrections. These corrections turn out to be small for MgII except for the features at $\lambda\lambda$ 4481 and 7877-96 \AA. For Vega (A0V), they found a correction of -0.21 dex for the 4481 \AA \ triplet which confirms our finding that this line systematically yields higher abundances.

\begin{figure*}
\centering
\vskip0.5cm
\includegraphics[scale=0.8]{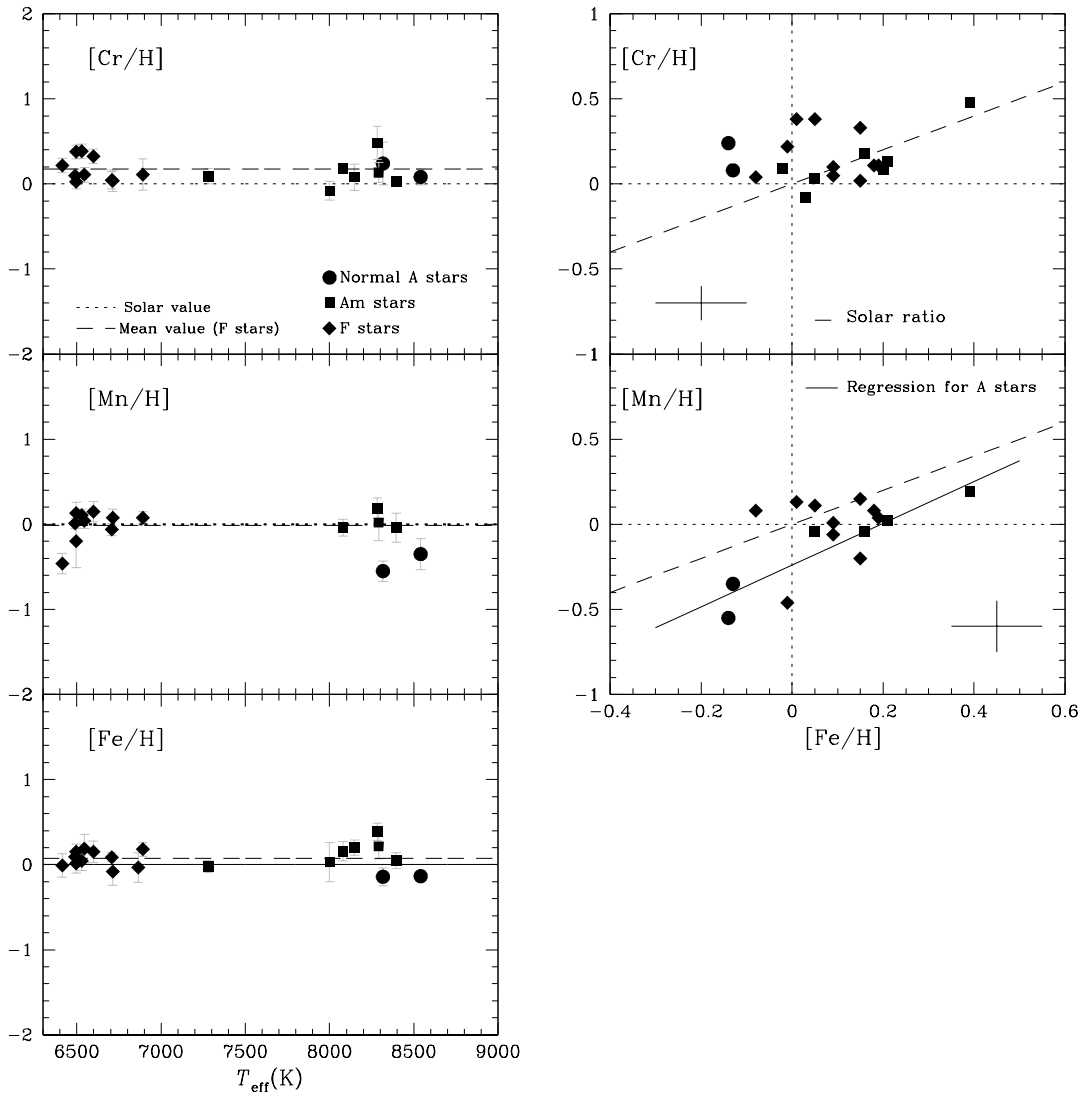}
\caption{Left panel: Abundance of chromium, manganese and iron versus effective temperature. The dotted line represents the solar value and the dashed one represents the mean abundance of F stars. Right panel: [Cr/H] and [Mn/H] versus [Fe/H]. The symbols are the same as in Figure \ref{CCA}. The dashed lines represent the solar ratios.}
\label{CrMnFe}
\end{figure*}

\subsubsection*{Silicon}

Nineteen lines of Si II of quality C to E have been synthesized. For the F stars, the silicon abundances hardly show any scatter around their mean value, +0.18 dex, slightly above solar. Almost all A stars are also overabundant in silicon and deviate very little from this mean value except for the Am star HD 107168 (Figure \ref{MgSiCa}). Thus there does not seem to be significant star-to-star variation in $[\frac{Si}{H}]$.\\
The silicon abundance does not show any convincing correlation with the iron abundance. Most of our data fall slightly above the line representing a solar silicon to iron ratio.
Note that Hill \& Landstreet (1993) did find a tight correlation between $[\frac{Si}{H}]$ and $[\frac{Fe}{H}]$. All their data fall close to the line representing a solar silicon to iron ratio. The lines we analysed differ from theirs and lead to overabundances possibly because of incorrect oscillator strengths or non-LTE effects.

\subsubsection*{Calcium}

Twelve lines of Ca II of quality C and D were used to derive the calcium abundance for all A and Am stars and for only two F stars.
The abundance of the F stars is subsolar ($<[\frac{Ca}{H}]>\sim-0.2 \ dex$). The two normal A stars are slightly deficient in calcium and the Am stars show modest deviations (both over and underabundances) around the solar abundance. Their maximum spread in [Ca/H] is 0.50 dex which is only marginally significant compared to the maximum estimated uncertainty (about 0.20 dex).\\
In a graph displaying the calcium abundance versus that of iron (Figure \ref{MgSiCa}), the Am stars, as expected, can easily be discriminated from the normal A and F stars. They all fall to the right in a region characterized by overabundances of iron and they are either Ca-deficient or Ca-rich. For the normal A and F stars, $[\frac{Ca}{H}]$ has a fairly uniform value (-0.20) dex and does not appear to vary with $[\frac{Fe}{H}]$.

\begin{figure*}
\centering
\vskip0.5cm
\includegraphics[scale=0.8]{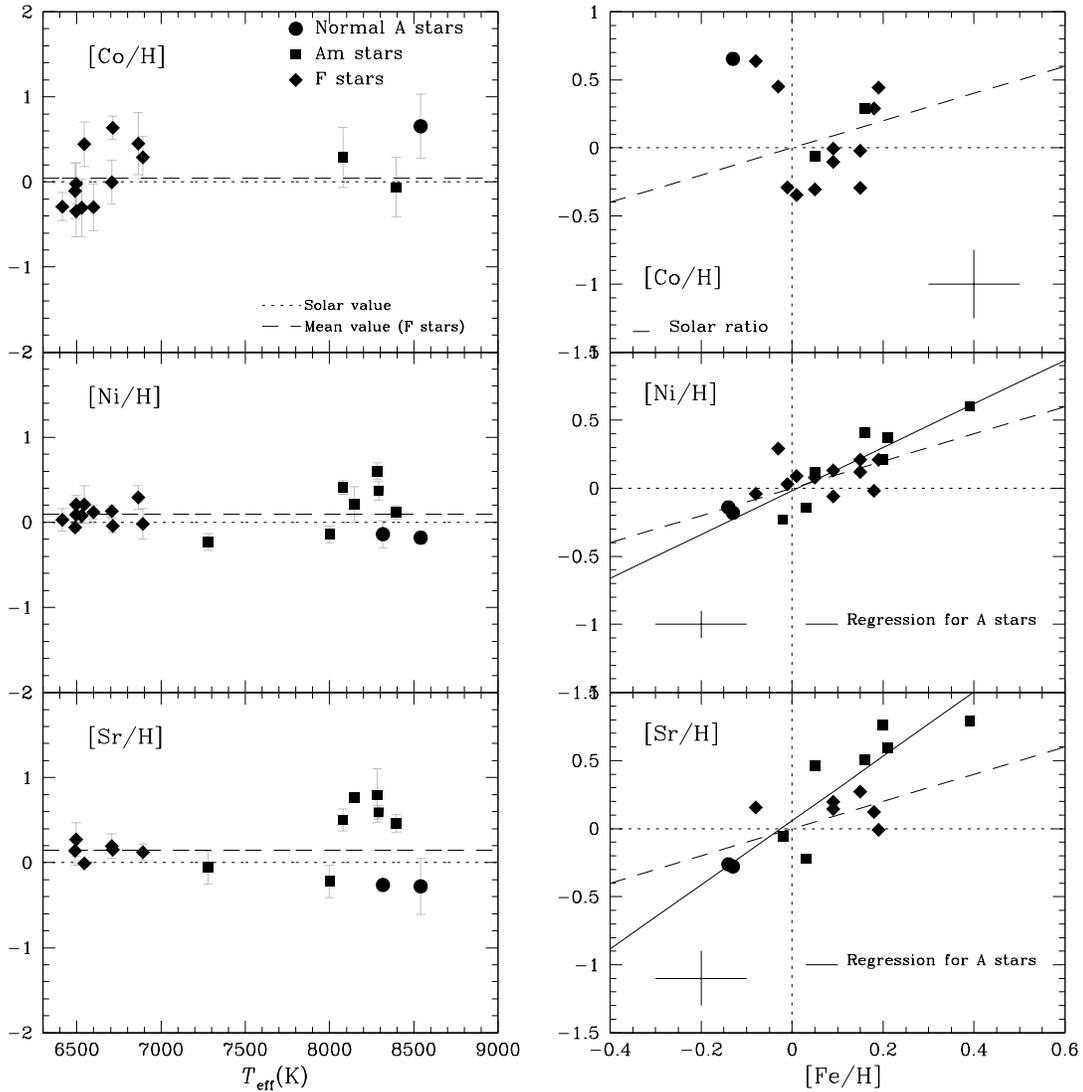}
\caption{Left panel: Abundance of cobalt, nickel and strontium versus effective temperature. The dotted line represents the solar value and the dashed one represents the mean abundance of F stars. Right panel: [Co/H], [Ni/Fe] and [Sr/H] versus [Fe/H]. The symbols are the same as in Figure \ref{CCA}. The dashed lines represent the solar ratios.}
\label{CoNiSr}
\end{figure*}

\subsubsection*{Scandium}

Eleven lines of quality D of Sc II were used to derive the scandium abundance. The scandium abundances of F stars are scattered very little  around the solar value.
About half of the A and Am stars are deficient in scandium (3 Am stars are close to solar while 4 exhibit deficiencies ranging from -0.2 dex to -1.2 dex).
The total spread in $[\frac{Sc}{H}]$ for these stars is about 1.20 dex, significantly larger than the typical uncertainty of 0.10 dex. There thus seems to be real star-to-star variation in $[\frac{Sc}{H}]$.\\
As for calcium, scandium does not exhibit any clear correlation or anticorrelation with respect to iron in the diagram of $[\frac{Sc}{H}]$ versus $[\frac{Fe}{H}]$ (Figure \ref{ScTiV}). The scandium abundances of the F and normal A stars, which are all very close to solar, do not depend on $[\frac{Fe}{H}]$. Most Am stars lie in the lower right part of the diagram: they are all iron-rich and 4 out of 7 are deficient in scandium.

\begin{figure*}
\centering
\vskip0.5cm
\includegraphics[scale=0.8]{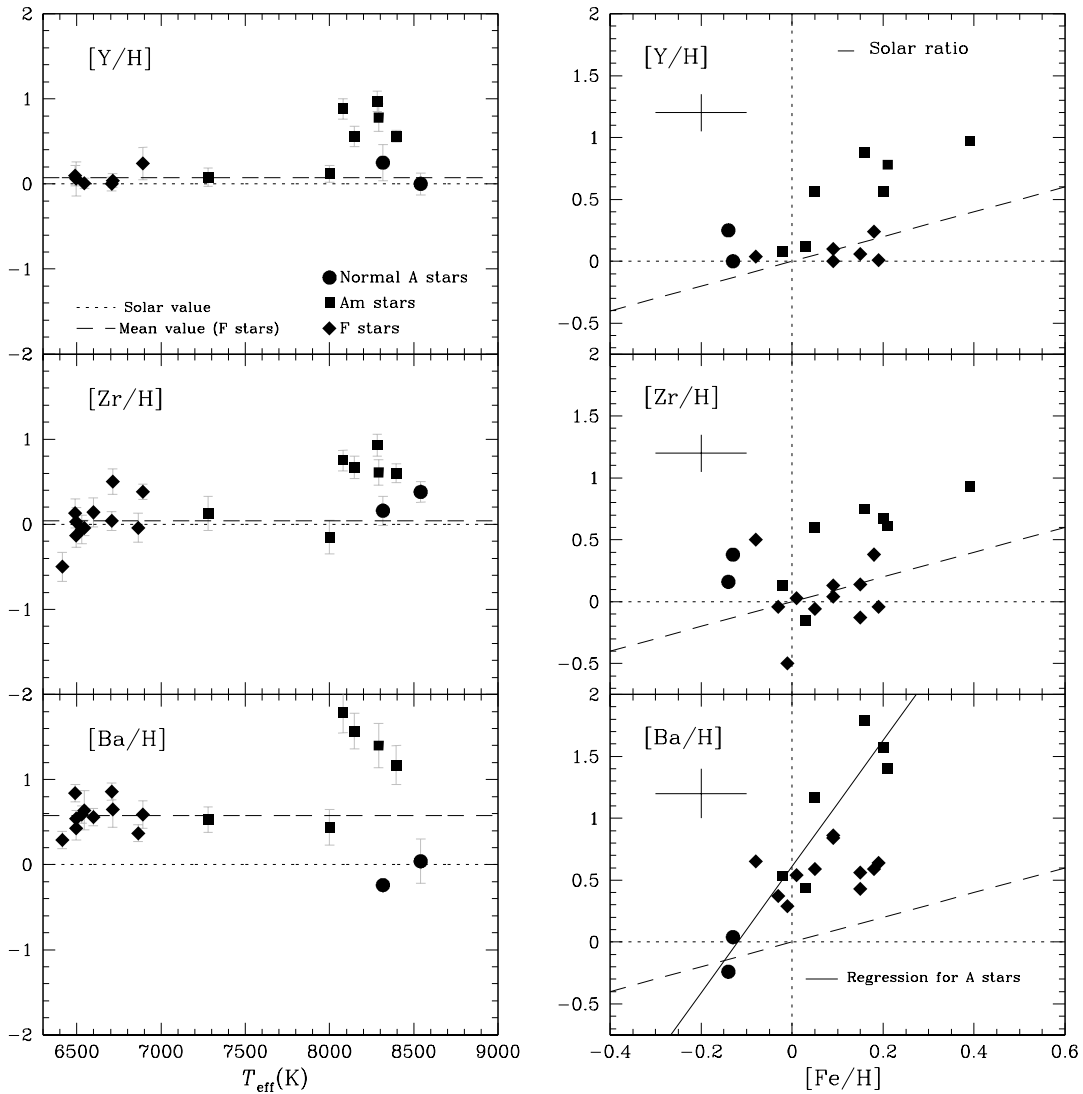}
\caption{Left panel: Abundance of yttrium, zirconium and barium versus effective temperature. The dotted line represents the solar value and the dashed one represents the mean abundance of F stars. Right panel: [Y/H], [Zr/Fe] and [Ba/H] versus [Fe/H]. The symbols are the same as in Figure \ref{CCA}. The dashed lines represent the solar ratios.}
\label{YZrBa}
\end{figure*}

\subsubsection{The iron peak elements}

\subsubsection*{Titanium, vanadium, chromium and manganese}

Twenty six lines of Ti II, most of them of quality D, have been synthesized. The oscillator strengths for this element are not as secure as for iron and chromium. One should not expect these abundances to be too reliable on an absolute scale.
The F stars show little scatter around their mean $[\frac{Ti}{H}]$ value, about +0.12 dex, slightly above solar. The A and Am stars are more scattered around this mean value. The spread in $[\frac{Ti}{H}]$ is about 0.68 dex which is larger than the typical uncertainty of about 0.08 dex. There seems to be real star-to-star variation for this element (Figure \ref{ScTiV}).
The titanium abundance appears to be loosely correlated to that of iron (correlation coefficient 0.66) for normal F and A stars. The $[\frac{Ti}{Fe}]$ ratios of these normal stars appear to be close to or slightly higher than solar as found by Lemke (1989) for a sample of 16 normal A stars. Hill \& Landstreet (1993) found a strong correlation of [Ti/H] with [Fe/H](correlation coefficient 0.95), their 
$[\frac{Ti}{Fe}]$ ratios being slightly above the solar value.
\\
Nine lines of V II were synthesized whose uncertainties are not specified. These lines are intrinsically weak in all spectra. The abundances derived for this element must be taken with caution. The vanadium abundances for the F stars are fairly scattered around their mean value, +0.50 dex. Vanadium was not detected in the two normal A stars and is found to have large overabundances for 4 Am stars.
There is no convincing correlation between  $[\frac{V}{H}]$ and $[\frac{Fe}{H}]$ (Figure \ref{ScTiV}). Most of the data fot the normal A stars and the F stars fall above the line representing the solar $[\frac{V}{Fe}]$ ratio which may be due to poorly determined oscillator strengths. Hill \& Landstreet (1993) found a correlation of $[\frac{V}{H}]$ with $[\frac{Fe}{H}]$ using lines other than ours; their ratios $[\frac{V}{Fe}]$ are also significantly above solar.
\\
Eleven lines of Cr II of quality D have been synthesized. As for titanium, the A stars are only a little more scattered than the F stars. The total spread in 
$[\frac{Cr}{H}]$ for the A and Am stars is about 0.56 dex (0.36 dex for the F stars) while the maximum uncertainty on $[\frac{Cr}{H}]$ is about 0.20. Should the data for HD107168 be removed, the spread in $[\frac{Cr}{H}]$ drops to 0.30 dex and is not significant. The evidence for star-to-star variations in $[\frac{Cr}{H}]$ is therefore rather weak.
The chromium abundance appears to be loosely correlated with that of iron (correlation factor = 0.85). Most of the data fot the normal A stars and the F stars fall above the line representing the solar $[\frac{Cr}{Fe}]$ ratio (Figure \ref{CrMnFe}). Hill \& Landstreet (1993) also found a correlation of $[\frac{Cr}{H}]$ with $[\frac{Fe}{H}]$, their ratios $[\frac{Cr}{Fe}]$ being only marginally above solar.
\\
Twenty lines of Mn I of quality B to C+ were synthesized. The F stars show little scatter around the solar value. The total spread for the A and Am stars is about 0.6 dex which is larger than the maximum uncertainty of 0.2 dex, suggesting real star-to-star variation. The two normal A stars are deficient in manganese. The manganese abundance appears to be well correlated with that of iron (correlation factor = 0.94). Most of the ratios $[\frac{Mn}{Fe}]$ run parallel to and below the line representing the solar ratio (Figure \ref{CrMnFe}). Hill \& Landstreet (1993) found a correlation of $[\frac{Mn}{H}]$ with $[\frac{Fe}{H}]$ using lines other than ours; their ratios $[\frac{Mn}{Fe}]$ are only
marginally above solar.

\subsubsection*{Iron, cobalt and nickel:}

Twenty seven lines of Fe II of quality C to E were synthesized.
In a graph $[\frac{Fe}{H}]$ versus $T_{\rm{eff}}$ (Figure \ref{CrMnFe}), the F stars are only slightly scattered around their mean value, +0.07 dex, which is slightly higher than 
\cite{FB92}. These authors found $<[\frac{Fe}{H}]>=-0.05\pm0.03 \ dex$, based on the analysis of the equivalent widths of a few Fe I lines for 14 F stars of the cluster. Our usage of different techniques (model atmospheres and line synthesis) and different lines (Fe II) probably accounts for the difference in iron abundance. The total spread in $[\frac{Fe}{H}]$ for the A and Am stars is about 0.53 dex which is larger than the maximum estimated uncertainty (0.10 dex). There are thus real star-to-star variations in $[\frac{Fe}{H}]$.
\\
For cobalt, our analysis is based on 12 lines of Co I, whose errors in the oscillator strengths are unknown. Most of these lines are weak and often are blended with lines whose atomic parameters are not necessarily accurately known. We therefore do not expect these abundances to be reliable.
In a graph of $[\frac{Co}{H}]$ versus $T_{\rm{eff}}$ (Figure \ref{CoNiSr}), the cobalt abundances are much more scattered for the F stars than for any other chemical element; this scatter is probably largely due to blending species. There is only one data point for the normal A stars. We feel that the abundances are not reliable enough to claim to find real star-to-star variations in $[\frac{Co}{H}]$ and not to find a correlation with $[\frac{Fe}{H}]$. 
\\
Fifty six lines of Ni I of quality C+ to D have been synthesized. In the graph $[\frac{Ni}{H}]$ versus $T_{\rm{eff}}$,  nickel behaves in a similar manner as iron (Figure \ref{CoNiSr}).
The F stars are fairly well grouped around their mean abundance (+ 0.09 dex), their maximum spread being 0.25 dex. The A and Am stars scatter over 0.83 dex whereas their maximum uncertainty is about 0.20 dex. There seems to be real star-to-star variations for $[\frac{Ni}{H}]$. The nickel abundance is tightly correlated with that of iron (correlation coefficient 0.92). Most of the ratios $[\frac{Ni}{Fe}]$ are close to solar.

\subsubsection{The heavy elements}

\subsubsection*{Strontium, yttrium, zirconium and barium:}

Our study of strontium is based on only two lines at 4077.71 \AA \ and 4215.52 \AA\ whose accuracies are not specified in NIST. Errors in gf values may result in a considerable zero point shift with respect to the Sun. We therefore do not attach too much significance to apparently large absolute over-or underabundances.
The F star strontium abundances show little scatter around their mean value, + 0.15 dex. The A and Am stars display a much larger spread in abundances, about 1.07 dex, significantly larger than the maximum uncertainty (0.30 dex) (Figure \ref{CoNiSr}).
The strontium abundance is correlated quite closely to the iron abundance (correlation coefficient 0.90). Strontium abundances vary more rapidly than those of iron as found by Lemke (1990).\\
Five lines of yttrium whose accuracies are unknown were synthesized. The same holds for zirconium.  Yttrium and zirconium are found to be nearly solar in F stars (Figure \ref{YZrBa}). These elements are overabundant by about 0.50 dex for 6 of the Am stars and by about 0.40 dex in the normal A stars. The A and Am stars display fairly large spreads in abundances of Y and Zr, about 1.0 dex, much larger than the associated uncertainties, indicating real star-to-star variations in $[\frac{Y}{H}]$ and $[\frac{Zr}{H}]$.\\
Five lines of quality B of barium have been synthesized. This element is found to be overabundant in all A, Am and F stars by large amounts (up to 1.10 dex).
The spread for $[\frac{Ba}{H}]$ is about 1.8 dex, much larger than the typical uncertainties, indicating real star-to-star variations in $[\frac{Ba}{H}]$ (Figure \ref{YZrBa}).
However, these overabundances may reflect a non-LTE effect, namely an overionization of barium in A stars. In Vega, the non-LTE Ba II abundances are lower than the LTE abundances by about 0.30 dex as demonstrated by \cite{1988A&A...192..264G} and \cite{1990A&A...240..331L} using the KIEL code. Non-LTE corrections for Ba should be sensitive to [Fe/H] which can differ much from star to star. Detailed calculations for the respective fundamental parameters and iron abundances of each star should therefore be performed . \\

\begin{table}
 \caption{Mean abundances and their respective dispersion for Fe, C, O, Ni, Sc, Si, Ba, Y, Zr and Sr for the F, normal A and Am stars.}
 \large
 \centering
 \label{abon-moyenne-deviation}
\begin{tabular}{c|cc|cc}
\hline
Elements&F stars&$\sigma_{F}$ &  A stars &  $\sigma_{A}$ \\   \hline 
$[\frac{C}{H}]$		&-0.01	&0.06	&-0.48	&0.24	   \\  \hline
$[\frac{O}{H}]$		&-0.11	&0.18	&-0.34	&0.31	   \\  \hline
$[\frac{Si}{H}]$	&0.18	&0.07	&0.22	&0.20	   \\  \hline
$[\frac{Sc}{H}]$	&-0.02	&0.04	&-0.23	&0.38	    \\ \hline
$[\frac{Fe}{H}]$	&0.07	&0.09	&-0.14	&0.16	          \\   \hline 
$[\frac{Ni}{H}]$	&0.09	&0.11	&0.11	&0.28	   \\  \hline
$[\frac{Sr}{H}]$	&0.15	&0.08	&0.25	&0.43	    \\ \hline
$[\frac{Y}{H}]$		&0.07	&0.08	&0.47	&0.34	   \\  \hline
$[\frac{Zr}{H}]$	&0.04	&0.34	&0.46	&0.33	      \\ \hline
$[\frac{Ba}{H}]$	&0.58	&0.17	&0.84	&0.70	   \\  \hline

\end{tabular}
   \end{table}


\subsection{Microturbulent velocities}

\begin{figure}
   \centering
  \includegraphics[width=9cm,height=7.5cm]{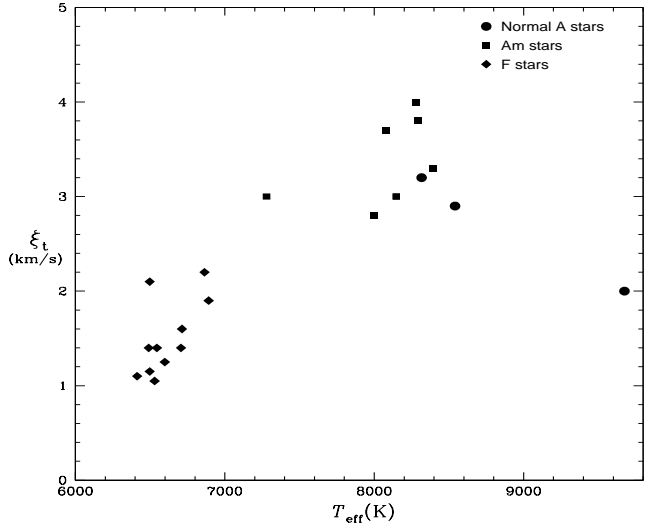}
      \caption{Variation of the derived microturbulence velocities with effective temperature.}
         \label{vmicr-teff}
	 \end{figure}

 A byproduct of this analysis has been to determine microturbulent velocities for each star. Figure \ref{vmicr-teff} displays the derived $\xi_{t}$ versus effective temperature. The overall variation agrees well with that found by Coupry \& Burkhart (1992) who found that $\xi_{t}$ varies from 0  km$\cdot$s$^{-1}$  for late B-type stars, up to about 3 km$\cdot$s$^{-1}$ for mid-A type stars down to around 2 km$\cdot$s$^{-1}$ for early F-type stars. Gray (2001) also found that $\xi_{t}$  varies diminishes 3 km$\cdot$s$^{-1}$ for mid-A type stars to about 1 km$\cdot$s$^{-1}$ for solar-type stars.

\subsection{Rotational velocities}
\begin{figure}
\centering
\includegraphics[width=9cm,height=7.5cm]{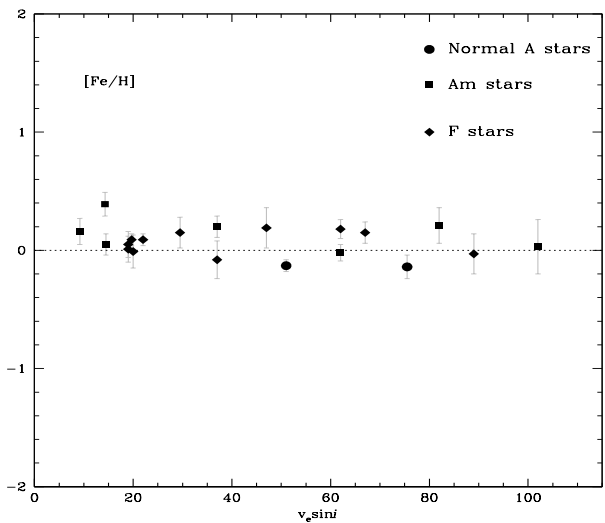}
\caption{Abundance of iron versus $v_{e}\sin i$ for A,Am and F stars.}
\label{Fe-vsini}
\end{figure}

The inferred rotational velicities $v_{e}\sin i$ are lower than 102 km$\cdot$s$^{-1}$. None of the found abundances appears to depend on $v_{e}\sin i$. The profiles of [X/H] versus $v_{e}\sin i$ are flat as shown for instance for iron in Figure \ref{Fe-vsini}. This result is not surprising. Detailed calculations of diffusion in the presence of meridional circulation carried out by \cite{1991ApJ...370..693C} revealed that, in stars rotating at less than $v_{e}\sin i$= 100 km$\cdot$s$^{-1}$, meridional circulation has little influence on chemical separation once the helium superficial convective zone has disappeared. Accordingly, the abundances should not present any positive nor negative trend (slope) with $v_{e}\sin i$ in this velocity regime. The absence of fast rotators in Coma Berenices (ie. $v_{e}\sin i>$ 100 km$\cdot$s$^{-1}$) prevents us from investigating the behaviour of the various abundances with rotational velocity above that velocity.

\section{Astrophysical implications}


The most important result of this study is the evidence of large star-to-star abundance variations for A stars in Coma Berenices. These stars appear to display much larger star-to-star variations in their abundances than the F stars do for the following chemical elements: C, O, Na, Sc, Ti, Mn, Fe, Ni, Sr, Y, Zr and Ba. In contrast, the abundances of Mg, Si, Ca and Cr do not show significant star-to-star variations.
For the Hyades, Varenne \& Monier (1999) found a similar behaviour for the abundances of C, O, Na, Sc, Fe and Ni. Monier (2005) also found large star-to-star variations in [Fe/H], [Ni/H] and [Si/H] for several A stars of the Uma group.\\
 \textit{We theorize that this peculiar behaviour is a signature of the occurence of transport processes competing with radiative diffusion (eg. rotational mixing in radiative zone, Zahn 2005)}. Indeed, if radiative diffusion was the only process at work, the microscopic diffusion velocity of a given chemical element should be the same in stars of similar effective temperatures and surface gravities. We would therefore expect similar surface compositions for stars of similar fundamental parameters.

\subsection{Self consistent modelling}

We have compared the derived abundances to the predictions of recent evolutionary models, calculated with the Montr\'eal code using slightly different assumptions and physics. This code treats radiative diffusion in detail and allows the inclusion of turbulent diffusion. \cite{1969A&A.....3..331S} proposed two possible physical origins of turbulent diffusion: loss of angular momentum while the star descends towards the Main Sequence or meridional circulation.

\subsubsection{The F-type stars}

For the F stars, the found abundances have been compared to the predictions of Turcotte et al's (1998) evolutionary models at the age of Coma Berenices.
Their models predict the evolution of the abundances for an F star
 of a given mass consistently with the internal structure.  They include the effects of gravitational sedimentation and of radiative diffusion for 28 chemical elements ($Z \leq 28$) but no macroscopic mixing (wind, accretion, meridional circulation, turbulence). These models are relevant for
F stars with masses in the range 1.1$M_\odot$ to 1.5$M_\odot$ from the pre-main sequence state up to hydrogen core exhaustion. The impact of the abundance variations with time on the structure of the star is taken into account. Monochromatic OPAL opacities for each element are used to
recalculate the opacity corresponding to the abundances and local conditions during the evolution. 
Atomic diffusion has an important effect on the opacities for stars more massive than $1.3 M_{\odot}$. In
these objects, the abundances of Fe and other iron-peak elements were found to substantially vary with time.  
\\
For the light elements C and O, the predicted large underabundances by Turcotte et al. (1998) for stars with $T_{eff} > $ 6500 K are not observed in our data. For these elements we find that the observed abundances in F stars are rather constant with a 
solar value for C
($<[\frac{C}{H}]> = -0.01\pm$0.06 dex) and marginally underabundant for O 
($<[\frac{O}{H}]> = -0.11 \pm$0.18 dex). In contrast, the predicted solar Na abundances for stars with $ 5950 \leq T_{eff} \leq$ 6700 K match well the observed solar abundances of F stars. For the F stars of 1.4$M_\odot$, the predicted Mg and Si underabundances are not seen either in our data. For the iron peak elements Fe and Ni, the predicted overabundances due to microscopic diffusion are not observed either.

\subsubsection{The A stars}

For A stars, we have compared the found abundances with the predictions of the recent models of  \cite{2000ApJ...529..338R}. In these new models, the effect of atomic and turbulent diffusion were calculated for stars of 1.45-3 $M_{\odot}$. They showed that the superficial abundances of the 28 species calculated in their models depend, in a star of a given mass, on essentially the initial metallicity and the deph of the zone mixed by turbulence. Figures 10 and 11 from \cite{2000ApJ...529..338R} represent the variations of surface abundances with time of the 28 elements for stars of 2-2.5-3 $M_{\odot}$. For the 2 $M_{\odot}$ model, \cite{2000ApJ...529..338R} found that the ratio of the abundances of carbon and oxygen at 450 Myr to their initial values decrease with time. We observed the same trend for C and O in Am stars using the mean abundances of  C and O in the F stars as initial abundances of these elements in the cluster.\\

\begin{figure}
   \centering
  \includegraphics[scale=0.55]{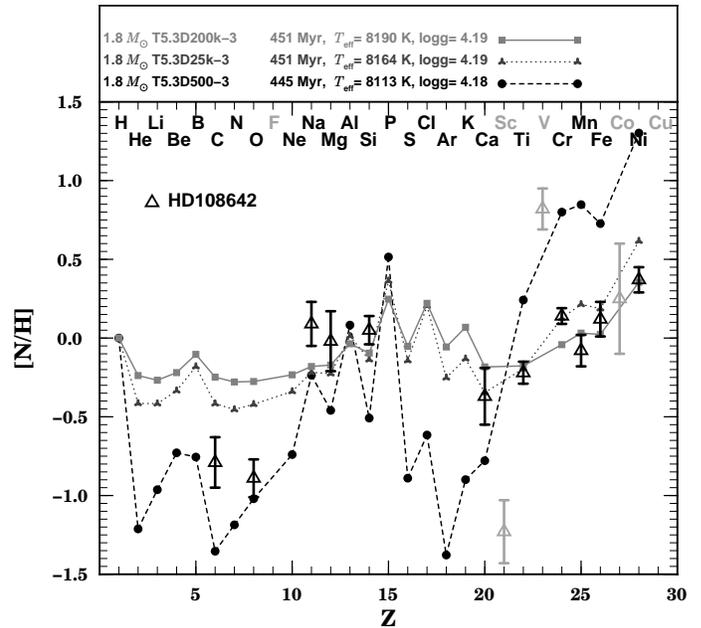}
      \caption{Comparison of the observed abundance pattern of HD108642 (A2m) with the predictions of three models calculated for a mass of 1.8 $M_{\odot}$ and different amounts of turbulent diffusion. Observed abundances are represented as triangles with error bars. Note that the models do not predict the surface abundances of Sc, V and Co.}
         \label{modele-evolution-HD108642}
	 \end{figure}

\begin{figure}
   \centering
  \includegraphics[scale=0.55]{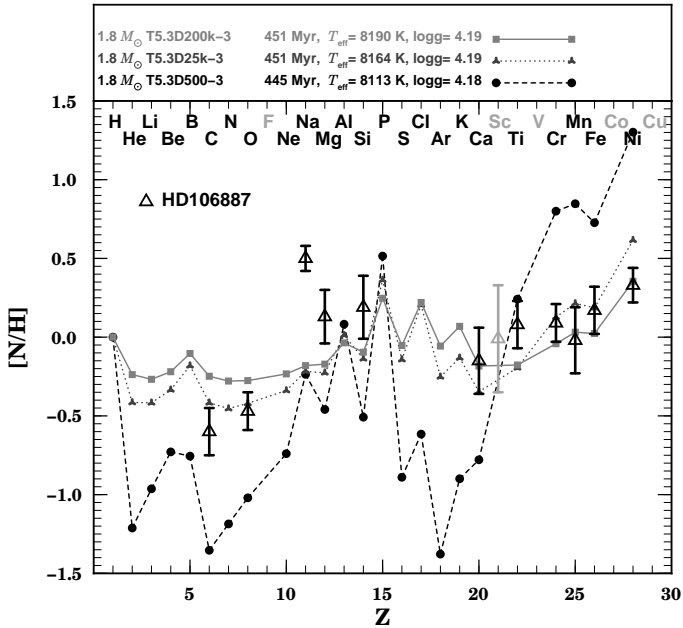}
      \caption{Comparison of the observed abundance pattern of HD106887 (A4m) with the same models.}
         \label{modele-evolution-HD106887}
	 \end{figure}

In addition, we have specifically computed a series of models for two Am stars of similar effective temperatures but different rotational velocities: HD108642, a slow rotator ($v_{e}\sin i$ = 9 km$\cdot$s$^{-1}$) and HD106887, a faster rotator ($v_{e}\sin i$ = 82 km$\cdot$s$^{-1}$). Masses for these two stars should be in the range 1.8 to 2.0 $M_{\odot}$. Seven models for a 1.8 $M_{\odot}$ mass star having different turbulent coefficients, $D_{T}$ (equation 1 in Richer et al. 2000) at the age of the Coma were computed. 
 The adopted initial and homogeneous abundances for these models are collected in Table \ref{tab:Xinit}. The models use an Eggleton-Faulkner-Flannery equation of state (\cite{EgFaFl73}) including Coulomb correction on the pressure (labeled as CEFF models) (see also \cite{ChristensenDalsgaardetal92}). The nuclear energy generation follows the prescriptions of \cite{BahcallPi92a}. These models take into account gravitational settling, thermal diffusion and radiative accelerations. The detailed treatment of atomic diffusion is described in \cite{TuRiMiIgRo98} and the radiative accelerations are from \cite{TuRiMiIgRo98} with correction for redistribution from \cite{GoLeArMi95} and \cite{LeMiRi2000}. These models are self consistent as the Rosseland opacity and radiative accelerations are recomputed at each time step in each layer for the exact local chemical composition using OPAL monochromatic opacities for 24 elements. Convection and semi-convection are modeled as diffusion processes as described in \cite{2000ApJ...529..338R} and \cite{RiMiRi2001}. The initial metallicity is taken from \cite{FB92}. \\
The models are compared to the abundance patterns of HD108642 in Figure \ref{modele-evolution-HD108642} and for HD106887 in Figure \ref{modele-evolution-HD106887}. None of the models reproduces entirely the characteristic abundance pattern, ie. marked underabundances of light elements and overabundances of iron-peak elements. They basically differ by the amount of turbulent diffusion included. The model that best approaches the abundance patterns for elements with Z$>$20 for these two stars is the model labeled as 1.8T5.3D200K-3 following the syntax of Table 1 of \cite{2000ApJ...529..338R}. In this model, the turbulent diffusion coefficient $D_{T}$ varies with density as:
\begin{center}

$D_{\rm{T}}=\omega D(\rm{He})_{0}\left(\frac{\rho_{0}}{\rho}\right)^{\it{n}}$
\end{center}
where n=3, $D(\rm{He})_{0}$ is the atomic diffusion coefficient of He at the density $\rho_{0}=\rho(T_{0})$ in the iron convection zone (see equation 1 of Richer et al. 2000, here $\omega=200\cdot10^{3}$ and $\log T_{0}=5.3$).
However this model predicts abundances that are almost 0.8 dex too large for C and O but it comes close to the abundances of Na, Mg and Si and follows the overabundances of iron-peak and heavier elements. Conversely, models with less turbulent diffusion 1.8T5.3D25K-3 and 1.8T5.3D500-3 roughly account for the abundances of elements with Z$<$15 but predict too large overabundances of iron-peak elements. \\
Part of the discrepancy between observed and theoretical abundance patterns could be due to non-LTE effects. Correcting the magnesium abundance for the magnesium triplet at $\lambda$4481 \AA \ reduces it by $\sim$0.2 dex and provides a better agreement with the T5.3D200K-3 model for this element. Lowering the sodium abundances, which are likely to be affected by non-LTE effects, will also improve the agreement with this model. However, the inclusion of competing processes in the models such as rotational mixing in the radiative zone for A stars, internal waves for F stars could also help improve the agreement.


\begin{table}
\caption{Initial chemical composition}
\begin{tabular}{lc}
\hline
Element & Mass fraction \\

H  \dotfill        & 7.03$\times 10^{-1}$ \\
$^{4}\rm{He}$ $^{(a)}$ \dotfill  & 2.7995$\times 10^{-1}$ \\
$^{12}$C $^{(b)}$ \dotfill & 2.935$\times 10^{-3}$ \\
N  \dotfill & 9.000$\times 10^{-4}$ \\
O  \dotfill & 8.189$\times 10^{-3}$ \\
Ne \dotfill & 1.675$\times 10^{-3}$ \\
Na \dotfill & 3.396$\times 10^{-5}$ \\
Mg \dotfill & 6.377$\times 10^{-4}$ \\
Al \dotfill & 5.519$\times 10^{-5}$ \\
Si \dotfill & 6.878$\times 10^{-4}$ \\
P  \dotfill & 5.944$\times 10^{-6}$ \\
S  \dotfill & 3.592$\times 10^{-4}$ \\
Cl \dotfill & 7.642$\times 10^{-6}$ \\
Ar \dotfill & 9.170$\times 10^{-5}$ \\
K  \dotfill & 3.396$\times 10^{-6}$ \\
Ca \dotfill & 6.368$\times 10^{-5}$ \\
Ti \dotfill & 3.396$\times 10^{-6}$ \\
Cr \dotfill & 1.698$\times 10^{-5}$ \\
Mn \dotfill & 9.340$\times 10^{-6}$ \\
Fe \dotfill & 1.219$\times 10^{-3}$ \\
Ni \dotfill & 7.557$\times 10^{-5}$ \\ \hline

 {(a)}  $^{3}He=5.000 \times 10^{-5}$ \\
 {(b)}  $^{13}$C is 1\% of $^{12}$C 
\end{tabular}
\label{tab:Xinit}
\end{table}

\section{Conclusion}


High and medium resolution spectra of 11 A and 11 F star members of the Coma Berenices open cluster have been synthesized in order to determine the abundances of C, O, Na, Mg, Si, Ca, Sc, Ti, V, Cr, Mn, Fe, Co, Ni, Sr, Y, Zr and Ba. 
In graphs representing the abundance [X/H] versus effective temperature, the A stars display abundances that are more scattered around the mean value than the F stars. Large star to star variations are detected for A stars for C, O, Na, Sc, Ti, Mn, Fe, Ni, Sr, Y, Zr and Ba which we interpret as evidence of transport processes competing with radiative diffusion. \\
The chemical pattern found for the A and F dwarfs of Coma resembles that found for the Hyades (Varenne \& Monier 1999) and the UMa group (Monier 2005). The mean iron abundance derived for the F stars is found to be  $<[\frac{Fe}{H}]>=0.07\pm0.09 \ dex$, slightly higher than the metallicity derived by \cite{FB92}. The abundances of manganese, nickel, strontium and barium are strongly correlated with the iron abundance for A and Am stars. The ratios [C/Fe] and [O/Fe] appear to be anticorrelated with [Fe/H]. The ratio [Ti/Fe] is solar as found by Lemke (1989). 
\\
The Am stars in Coma Berenices are found to be deficient in light elements (C and O), they are not all deficient in calcium and scandium but are all overabundant in metallic and heavy elements (Fe, Co, Ni, Sr, Y, Zr and Ba). The two normal A stars have almost solar abundances. The F stars have solar abundances for almost all the elements except for Mg, Si, V and Ba.\\
The abundance patterns predicted by current state of the art evolutionary models following the prescriptions of \cite{2000ApJ...529..338R} have been compared to the observed abundance patterns of two Am stars of the cluster, one a slow rotator (HD108642) and one a moderately fast rotator (HD106887). 
These models were calculated with different strengths of the turbulent diffusion coefficient. None of the models reproduces entirely the characteristic abundance pattern, ie. marked underabundances of light elements and overabundances of iron-peak and heavier elements. Part of the discrepancy may arise from non-LTE effects. In this respect, non-LTE abundance determinations for many of the elements analysed here (when feasible, ie. when atomic data and model atoms are available) are highly desirable. However it is likely that the inclusion of competing processes such as rotational mixing in the radiative zones (which will vary from star to star) should also help reproduce the observed abundance patterns and the large scatter of abundances of several elements in A stars.

\begin{appendix}
\section{Determination of uncertainties}
Six major sources are included in the uncertainty determinations: uncertainty on the effective temperature ($\sigma_{T_{\rm{eff}}}$), on the surface gravity ($\sigma_{\log g}$), on the microturbulent velocity ($\sigma_{\xi_{t}}$), on the apparent rotational velocity ($\sigma_{v_{e}\sin i}$), the oscillator strength ($\sigma_{\log gf}$) and the continuum placement ($\sigma_{cont}$). These uncertainties are supposed to be independent, so that the total uncertainty $\sigma_{tot_{i}}$ for a given transition (i) is:\\
\begin{equation}
\sigma_{tot_{i}}^{2}=\sigma_{T_{\rm{eff}}}^{2}+\sigma_{\log g}^{2}+\sigma_{\xi_{t}}^{2}+\sigma_{v_{e}\sin i}^{2}+\sigma_{\log gf}^{2}+\sigma_{cont}^{2}.
\end{equation}  
The mean abundance $<[\frac{X}{H}]>$ is then computed as a weighted mean of the individual abundances [X/H]$_{i}$ derived for each transition (i):\\
\begin{equation}
<[\frac{X}{H}]>=\frac{\sum_{i}([\frac{X}{H}]_{i}/\sigma_{tot_{i}}^{2})}{\sum_{i}(1/\sigma^{2}_{tot_{i}})}
\end{equation}
and the standard deviation, $\sigma_{sd}$ is given by: \\
\begin{equation}
\frac{1}{\sigma_{sd}^{2}}=\sum_{i=1}^{N}(1/\sigma_{tot_{i}}^{2})  
\end{equation}
where N is the number of lines per element. This procedure has been applied to 4 A stars (HD107966, HD107655, HD107168 and HD107513) and 3 F stars (HD106103, HD107611 and HD109530). In the fastest rotators, the pseudo-continuum, $f_{pseudo}^{\lambda}$, we see that regions free of lines in the observed spectra are actually affected by the rotational broadening of neighbouring lines. The level of the actual continuum, $f_{cont}^{\lambda}$, can be recovered by carefully synthesizing the pseudo-continuum windows, yielding the theoretical flux, $F_{pseudo}^{\lambda}$, and the actual corresponding continuum flux, $F_{cont}^{\lambda}$, at the appropriate velocity $v_{e}\sin i \pm \Delta(v_{e}\sin i)$. The observed intensity level in these pseudo-continuum windows was then multiplied by ($F_{cont}^{\lambda}$/$F_{pseudo}^{\lambda}$) to recover the observed continuum level: 

\begin{equation}
f_{cont}^{\lambda}=\frac{F_{cont}^{\lambda}}{F_{pseudo}^{\lambda}} f_{pseudo}^{\lambda}.
\end{equation}
The maximum and minimum allowed rotational velocity $v_{e}\sin i \pm \Delta(v_{e}\sin i)$ yields 2 ratios $\frac{F_{cont}^{\lambda}}{F_{pseudo}^{\lambda}}$($v_{e}\sin i_{max}$) and $\frac{F_{cont}^{\lambda}}{F_{pseudo}^{\lambda}}$($v_{e}\sin i_{min}$). The corresponding observed normalized line profiles were then used to derive the corresponding changes in abundances due to the different locations of the continuum. This test was performed in several spectral regions (excluding overlaping regions of 2 orders) and yields a maximum uncertainty of about 0.07 dex on the abundances.

 For the other stars, the final abundances are averages. The errors on the elemental abundances are standard deviation assuming a Gaussian distribution of the abundances derived from each line
\begin{equation}
\bar{x}=\frac{\sum_{i}^{}x_{i}}{N}
\end{equation}
\begin{equation}
\sigma^{2}=\frac{\sum_{i}^{}(x_{i}-\bar{x})^{2}}{N}
\end{equation}
where $\bar{x}$ is the mean value of the abundance, N the number of lines of the element and $\sigma$ the standard deviation.\\
\end{appendix}

\begin{acknowledgements}
We warmly thank the OHP night staff for the support during the observing runs. This research has used the SIMBAD, WEBDA, VALD, NIST and Kurucz databases.        
      
\end{acknowledgements}

\longtab{8}{
\centering
\begin{longtable}{lcccc|lcccc}
 
\caption{Linelist used for abundance determination. The "A" term is used for a $gf$ accurancy lower than 3\%, "B" lower than 10\%, "C+" lower than 18\%, "C" lower than 25\%, "D+" lower than 40 \%, "D" lower than 50 \% and "E" higher than 50 \%. If no accurancy is available, we used the E (\textit{i.e.} 50\%) value in the uncertainties calculations.  References are FMW for \cite{1988atpi.book.....F}, PTP for \cite{2002ApJS..138..247P}, NBS for \cite{1969AD......1....1M}, KO83 for \cite{1983AAfz...49...39K}, SL90 for \cite{1990MNRAS.247..611S}, BL48 for \cite{1948ZA.....25..325B}, Bal81 for \cite{1981ApJ...248..867B}, MC86 for \cite{1986ApJ...308..254M}, W85 for \cite{1985MNRAS.213...71W} and Kurucz (http://kurucz.harvard.edu/LINELISTS/GFALL/) for the gfall.dat linelist.}
\label{linelist}
\\ \hline
Element   &   $\lambda$(\AA)  &  $\log gf$  &      Accuracy  & references  & Element   &   $\lambda$(\AA)  &  $\log gf$  &      Accuracy  & references    \\ (Ionization) &&&&& (Ionization) &&&&\\ \hline
\endfirsthead
\caption{continued.}
 \\ \hline
 
Element   &   $\lambda$(\AA)  &  $\log gf$  &      Accuracy  & 
references  & Element   &   $\lambda$(\AA)  &  $\log gf$  &      Accuracy  & 
references    \\ 
(Ionization) &&&&& (Ionization) &&&&\\ \hline
\endhead
\hline 
\endfoot

\hline 
\endlastfoot
CI   &   4371,3670   &   -1,962   &   B   &   NIST     &      SiII &   4072,7090   &   -2,367	&	&  SL90  \\
CI   &   4932,0490   &   -1,658   &   B   &   NIST     &      SiII &   4075,4520   &   -1,403	&	&  SL90  \\
CI   &   5052,1670   &   -1,303   &   B   &   NIST     &      SiII &   4128,0540   &   0,306	&   C	&   NIST \\
CI   &   5380,3370   &   -1,616   &   B   &   NIST     &      SiII &   4130,8940   &   0,464	&   C	&   NIST \\
CI   &   5793,1200   &   -2,063   &   B   &   NIST     & 	SiII &   4190,7240   &   -0,351   &	  &   NIST		  \\	 
CI   &   5800,6020   &   -2,337   &   B   &   NIST     &      	SiII &   4198,1330   &   -0,611   &	  &   NIST	  \\
CI   &   6587,6100   &   -1,003   &   B   &   NIST     &      	SiII &   4621,4180   &   -0,540   &   D   &   Kurucz	  \\
     &     	     &    	  &	  &	       &      	SiII &   4621,6960   &   -1,675   &   D   &   Kurucz	  \\  
OI   &   3947,2950   &   -2,095   &   B   &   NIST 	&     	SiII &   4621,7220   &   -0,387   &   D   &   Kurucz	 \\   
OI   &   3947,4810   &   -2,244   &   B   &   NIST 	&     	SiII &   5041,0240   &   0,174    &   D+  &   NIST	  \\
OI   &   3947,5860   &   -2,467   &   B   &   NIST 	&     	SiII &   5055,9840   &   0,441    &   D+  &   NIST	  \\
OI   &   3947,9530   &   -1,761   &   B   &   FMW  	&     	SiII &   5056,3170   &   -0,535   &   E   &   NIST	  \\
OI   &   4368,1930   &   -2,665   &   B   &   NIST 	 &	 SiII &   5466,4320   &   -0,190   &   D   &   Kurucz	   \\
OI   &   4368,2420   &   -1,964   &   B   &   NIST 	 &	 SiII &   5669,5630   &   0,266    &	   &   NIST	   \\
OI   &   4368,2580   &   -2,818   &   B   &   NIST 	 &	 SiII &   5688,8170   &   0,106    &	   &   NIST	   \\
OI   &   5329,6730   &   -2,063   &   C+   &   NIST	 &	 SiII &   5957,5590   &   -0,349   &   D   &   NIST	   \\
OI   &   5329,6810   &   -1,473   &   C+   &   NIST	 &	  SiII &   5978,9300   &   -0,061   &	D   &	NIST	    \\	    
OI   &   5329,6900   &   -1,268   &   C+   &   NIST	 &	  SiII &   6347,1100   &   0,230    &	C   &	NIST	    \\	    
OI   &   5330,7260   &   -2,416   &   C+   &   NIST	 &	  SiII &   6371,3710   &   -0,080   &	C   &	NIST	    \\	    
OI   &   5330,7350   &   -1,570   &   C+   &   NIST	 &	    &		       &       &	       &	    \\	    
OI   &   5330,7410   &   -0,983   &   C+   &   NIST	 &     CaII   &   3933,6630   &   -0,135   &   C   &   NIST	\\
OI   &   6155,9610   &   -1,363   &   B   &   NIST 	 &     CaII   &   3968,4690   &   -0,180   &   C   &   NIST	\\
OI   &   6155,9710   &   -1,011   &   B   &   NIST 	 &     CaII   &   4472,0500   &   -2,694   &	  &   Kurucz	\\
OI   &   6155,9890   &   -1,120   &   B   &   NIST 	 &     CaII   &   4479,4330   &   -2,994   &	  &   Kurucz	\\
OI   &   6156,7370   &   -1,487   &   B   &   NIST 	 &     CaII   &   4489,1790   &   -0,726   &	  &   Kurucz	\\
OI   &   6156,7550   &   -0,898   &   B   &   NIST 	 &     CaII   &   4489,1790   &   -2,157   &	  &   Kurucz	\\
OI   &   6156,7780   &   -0,694   &   B   &   NIST 	 &     CaII   &   4489,1790   &   -0,613   &	  &   Kurucz	\\  
OI   &   6158,1490   &   -1,841   &   B   &   NIST 	 &     CaII   &   5001,4790   &   -0,517   &   D   &   NIST	  \\
OI   &   6158,1720   &   -0,995   &   B   &   NIST 	 &     CaII   &   5019,9710   &   -0,257   &   D   &   NIST	   \\   		OI   &   6158,1870   &   -0,409   &   B   &   NIST 	 &     CaII   &   5021,1380   &   -1,217   &   D   &   NIST	   \\
     &     	     &    	  &	  &		&      CaII   &   5285,2660   &   -1,153   &   D   &   NIST	   \\      
NaI  &	4494.1800    &	-1.840	  &	C &	NIST    &	CaII   &   5307,2240   &   -0,853   &   D   &   NIST	   \\
NaI  &	4497.6570    &	-1.574	  &	B &	NIST    &	    &		       &       &	       &	    \\	    
NaI  &	4668.5590    &	-1.310	  &	C &	NIST    &       ScII   &   4246,8220   &   0,242   &   D   &   NIST	\\
NaI  &	4978.5410    &	-1.210	  &	C &	NIST    &      ScII   &   4314,0830   &   -0,100   &   D   &   NIST    \\
NaI  &	4982.8130    &	-0.961	  &	C &	NIST    &      ScII   &   4320,7320   &   -0,250   &   D   &   NIST    \\
NaI  &	5889.9500    &	0.112	  &	A &	NIST    &      ScII   &   4324,9960   &   -0,440   &   D   &   NIST    \\
NaI  &	5895.9240    &	-0.191	  &	A &	NIST    &      ScII   &   4374,4570   &   -0,418   &   D   &   NIST    \\
NaI  &	6154.2260    &	-1.547	  &	A &	NIST    &      ScII   &   4670,4070   &   -0,576   &   D   &   NIST    \\
NaI  &	6160.7470    &	-1.230	  &	C &	NIST	&      ScII   &   5031,0210   &   -0,400   &   D   &   NIST    \\
     &     	     &    	  &	  &		&      ScII   &   5239,8130   &   -0,765   &   D   &   NIST    \\   			  
MgII &   4384,6370   &   -0,792   &	D &	NIST    &      ScII   &   5526,7900   &   0,020   &   D   &   NIST     \\
MgII &   4390,5140   &   -1,706   &	D &	NIST 	&      ScII   &   5657,8960   &   -0,603   &   D   &   NIST    \\
MgII &   4390,5720   &   -0,530   &	D &	NIST 	&      ScII   &   6604,6010   &   -1,310   &   D   &   NIST    \\ 
MgII &   4427,9940   &   -1,201   &	C+&   NIST	&         &		     &       &  	     &  	  \\  
MgII &   4481,1260   &   0,730    &   B   &  BL48	&	 TiII	&   4163,6440	&   -0,130   &   D   &   PTP	  \\
MgII &   4481,1500   &   -0,570   &	B &	BL48	&	 TiII	&   4287,8730	&   -1,790   &      &	PTP	  \\
MgII &   4481,3250   &   0,575    &   B   &  BL48	&	 TiII	&   4290,2190	&   -0,850   &      &	PTP	  \\
MgI  &   4702,9910   &   -0,374   &   C   &   NIST	&	 TiII	&   4294,0990	&   -0,930   &      &	PTP	  \\
MgI  &   5167,3213   &   -0,856   &   B   &   NIST	&	 TiII	&   4300,0420	&   -0,440   &   D   &   PTP	  \\
MgI  &   5172,6844   &   -0,380   &   B   &   NIST	&	 TiII	&   4316,7940	&   -1,420   &   D   &  KO83	  \\
MgI  &   5183,6034   &   -0,158   &   B   &   NIST	&	 TiII	&   4386,8440	&   -0,960   &      &	PTP	  \\
MgI  &   5528,4050   &   -0,498   &   B+  &  		&	 TiII	&   4394,0590	&   -1,780   &      &	PTP	  \\
     &     	     &    	  &	  &		&     	 TiII	&   4395,0310	&   -0,540   &   A   &   PTP	  \\
	 & 	  &  	  &  	   & 			   &	 TiII	&   4399,7720	&   -1,190   &      &	PTP	  \\
 TiII	&   4411,0720	&   -0,670   &   D   &   PTP	 &   FeII &   4233,1720   &   -2,000   &   C   &   FMW  	\\
 TiII	&   4417,7140	&   -1,190   &      &	PTP	 &  FeII &   4258,1540   &   -3,400   &   D   &   FMW		\\
 TiII	&   4443,8010	&   -0,720   &   D   &   PTP	 &  FeII &   4273,3260   &   -3,258   &   D   &   FMW		\\
 TiII	&   4468,4920	&   -0,600   &   D   &   FMW	 &  FeII &   4296,5720   &   -3,010   &   D   &   FMW		\\
 TiII	&   4488,3250	&   -0,510   &   D   &   PTP	 &  FeII &   4385,3870   &   -2,570   &   D   &   FMW		\\
 TiII	&   4501,2700	&   -0,770   &   D   &   PTP	 &  FeII &   4416,8300   &   -2,600   &   D   &   FMW		\\
 TiII	&   4549,6210	&   -0,470   &   D   &   PTP	 &  FeII &   4472,0900   &   -1,791   &       &   Kurucz	\\	  
 TiII &   4571,9710   &   -0,320   &   D   &   PTP	 &  FeII &   4472,6200   &   -2,340   &       &   Kurucz    \\
TiII &   4589,9580   &   -1,780   &   D   &   PTP	 &  FeII &   4472,9290   &   -3,430   &       &   Kurucz     \\  
TiII &   4629,2740   &   -2,240   &	  &   Kurucz	 &  FeII   &   4491,4050   &   -2,690	&   C	&   FMW       \\ 
TiII &   4657,2000   &   -2,240   &	  &   Kurucz	 &  FeII   &   4508,2880   &   -2,210	&   C	&   FMW       \\ 
TiII &   4805,0850   &   -1,100   &   D   &   PTP	 &  FeII   &   4515,3390   &   -2,490	&   C	&   FMW       \\ 
TiII &   5129,1560   &   -1,400   &   D   &  KO83	 &  FeII   &   4520,2240   &   -2,600	&   C	&   FMW       \\ 
TiII &   5188,6870   &   -1,050   &   D   &   PTP	&   FeII   &   4522,6340   &   -2,030	&   C	&   FMW       \\ 
TiII &   5336,7860   &   -1,590   &   A   &   PTP	&   FeII   &   4541,5240   &   -3,050	&   C	&   FMW       \\ 
       &               &	    &	&	    	&   FeII   &   4555,8900   &   -2,290	&	&   FMW       \\ 
VII  &  4475.6700    &  -1.440    &       & Kurucz   	 &  FeII   &   4576,3400   &   -3,040	&	&   FMW       \\ 	   
VII  &  4528.5000    &  -0.960    &       & Kurucz   	 &  FeII   &   4582,8350   &   -3,100	&   C	&   FMW       \\ 
VII  &  4532.1700    &  -0.760    &       & Kurucz   	 &  FeII   &   4620,5210   &   -3,280	&   D	&   FMW       \\ 
VII  &  4538.6200    &  -1.800    &       & Kurucz   	 &  FeII   &   4635,3160   &   -1,650	&   D	&   FMW       \\ 
VII  &  4558.4500    &  -0.930    &       & Kurucz   	 &  FeII   &   4656,9810   &   -3,630	&   E	&   FMW       \\ 
VII  &  4564.5900    &  -1.450    &       & Kurucz   	 &  FeII   &   4666,7580   &   -3,330	&   D	&   FMW       \\
VII  &  4577.1300    &  -2.140    &       & Kurucz   	 &  FeII   &   4923,9270   &   -1,320	&   C	&   FMW       \\
VII  &  4590.5000    & -0.780     &       & Kurucz   	 &  FeII   &   5197,5770   &   -2,100	&   C	&   FMW       \\
VII  &  4600.1800    & -1.360	  &       & Kurucz   	 &  FeII   &   5276,0020   &   -1,940	&   C	&   FMW       \\
     &     	     &    	  &	  &	     	 &  FeII   &   5316,6150   &   -1,850	&   C	&   FMW       \\
CrII &   4558,6500   &   -0,660   &   D   &   Kurucz 	 &  FeII   &   5506,1950   &   0,950	&   D	&   FMW       \\ 
CrII &   4588,1990   &   -0,643   &	  &   Kurucz 	 &  	   &		   &		&	&	     \\
CrII &   4592,0490   &   -1,217   &   D   &   FMW    	 &   CoI    &  4466.8800    &  -0.540	 &	 & Kurucz     \\
CrII &   4616,6290   &   -1,291   &	  &  SL90    	&    CoI    &  4469.5400    &  -0.330	 &	 & Kurucz      \\
CrII &   4618,8030   &   -1,110   &   D   &   Kurucz 	&    CoI    &  4471.5400    &  -0.770	 &	 &Kurucz       \\
CrII &   4634,0700   &   -0,990   &	  &   Kurucz 	&    CoI    &  4530.9500    &  0.150	 &	 &Kurucz	\\
CrII &   4812,3370   &   -1,995   &   D   &   Kurucz 	 &   CoI    &  4533.9800    & -0.500	 &	 & Kurucz	\\
CrII &   5237,3290   &   -1,160   &   D   &  FMW     	 &   CoI    &  4549.6500    & -0.330	 &	 &Kurucz	\\
CrII &   5308,4400   &   -1,810   &   D   &   FMW   	&    CoI    &  4565.5800    & -0.220	 &	 &Kurucz	\\
CrII &   5313,5900   &   -1,650   &   D   &   FMW   	&    CoI    &  4581.5900    & -0.150	 &	 & Kurucz	\\
CrII &   5502,0670   &   -1,990   &   D   &   FMW   	&    CoI    &  4594.6300    & -0.080	 &	 & Kurucz	\\
     &     	     &    	  &	  &		&    CoI    &  4596.8900    & -0.010	 &	 & Kurucz      \\
MnI  &   4451,5860   &   0,278    &   B   &   FMW     &      CoI    &  4625.7600    & -0.370	  &	 & Kurucz     \\  
MnI  &   4453,0120   &   -0,490   &   C+  &   FMW     &      CoI    &  4629.3600    &  -0.190	 &	 & Kurucz     \\ 
MnI  &   4457,0440   &   -0,555   &   C+  &   FMW      &    	    &		    &		 &	 &	      \\
MnI  &   4458,2540   &   0,042    &   C+  &   FMW      &  NiI	  &  4468.4340    &  -1.642 &	   &   Kurucz	     \\  
MnI  &   4461,0790   &   -0,380   &   C+  &  FMW       &  NiI	  &  4470.4720    &  -0.310 & D    &   Kurucz	     \\  
MnI  &   4462,0310   &   0,320    &   C+  &   FMW      &  NiI	  &  4480.5610    &  -1.491 &	   &   Kurucz	     \\  
MnI  &   4464,6820   &   -0,104   &   B   &   FMW      &  NiI	  &  4490.0490    &  -2.108 &	   &   Kurucz	     \\  
MnI  &   4470,1440   &   -0,444   &   B   &  FMW       &  NiI	  &  4490.5250    &  -2.324 &	   &   Kurucz	     \\  
MnI  &   4472,8060   &   -0,583   &   B   &   FMW      &  NiI	  &  4512.9860    &  -1.470 & D    &   Kurucz	     \\  
MnI  &   4490,0900   &   -0,521   &   B   &   FMW      &  NiI	  &  4519.9790    &  -2.880 & D+   &   FMW	     \\  
MnI  &   4498,9020   &   -0,343   &   B   &   FMW      &  NiI	  &  4521.3220    &  -0.949 &	   &   Kurucz	     \\  
MnI  &   4502,2130   &   -0,344   &   B   &   FMW      &  NiI	  &  4523.6940    &  -1.305 &	   &   Kurucz	     \\  
MnI  &   4709,7120   &   -0,339   &   B   &   FMW      &  NiI	  &  4528.5260    &  -1.127 &	   &   Kurucz	     \\  
MnI  &   4739,0870   &   -0,490   &   B   &   FMW      &  NiI	  &  4542.2370    &  -1.308 &	   &   Kurucz	     \\  
MnI  &   4754,0420   &   -0,085   &   B   &   FMW      &  NiI	  &  4546.9200    &  -0.271 &	   &   Kurucz	     \\  
MnI  &   4761,5120   &   -0,138   &   B   &   FMW     &   NiI	  &  4551.2170    &  -0.880 & D    &   FMW	     \\  
MnI  &   4762,3670   &   0,426    &   B   &   FMW     &   NiI	  &  4559.9210    &  -1.737 &	   &   Kurucz	     \\  
MnI  &   4783,4270   &   0,042    &   B   &   FMW     &   NiI	  &  4572.0410    &  -0.536 &	   &   Kurucz	     \\  
MnI  &   4823,5240   &   0,144    &   B   &   FMW      &  NiI	  &  4588.4110    &  -0.745 &	   &   Kurucz	     \\  
MnI  &   5255,3260   &   -0,763   &   B   &   FMW      &  NiI	  &  4592.5220    &  -0.370 &	   &   Kurucz	     \\  
     &     	     &    	  &	  &	      &   NiI	  &  4596.3830    &  -0.704 &	   &   Kurucz	     \\ 
     &     	     &    	  &	  &	      & NiI	&  4600.3550	&  -0.610 &	 &   FMW    	   \\
NiI	&  4604.9820	&  -0.250 & D	 &   Kurucz 	  & SrII    &  4077.7090    & 0.151   &      & NIST		\\
NiI	&  4606.2190	&  -1.000 & D	 &  FMW     	  & SrII    &  4215.5200    & -0.169  &      & NIST 		\\
NiI	&  4609.9050	&  -0.580 &	 &   Kurucz 	  & 	    &		    &	      &      &      		\\
NiI	&  4617.8620	&  -0.525 &	 &   Kurucz 	  & YII     &  4883.6840    &  0.070  &      &   kurucz 	\\
NiI	&  4631.0170	&  -0.957 &	 &   Kurucz 	  & YII     &  4900.1200    &  -0.09  &      &   kurucz 	\\
NiI	&  4648.6460	&  -0.100 & D	 &   Kurucz 	  & YII     &  4982.1290    &  -1.290 &      &   kurucz 	\\
NiI	&  4886.7050	&  -1.780 &	 &   Kurucz 	  & YII     &  5087.4160    &  -0.170 &      &   kurucz 	\\
NiI     &  4886.9760    &  -1.120 &      &   Kurucz	  & YII     &  5200.4060    &  -0.570 &      &   kurucz 	\\
NiI     &  4900.9670    &  -1.670 & E    &   FMW  	  & 	    &		    &	      &      &      	       \\
NiI	&  4904.4070	&  -0.170 & D	 &   FMW	  & ZrII    &  4149.2170    &  -0.030 &      &   kurucz        \\
NiI	&  4912.0200	&  -0.800 & D	 &   FMW	  & ZrII    &  4156.2400    &  -0.776 &      &   kurucz        \\
NiI     &  4913.9680    &  -0.630 & D	 &   FMW          & ZrII    &  4161.2100    &  -0.720 &      &   kurucz        \\
NiI     &  4918.3620    &  -0.240 & D	 &   FMW    	  & ZrII    &  4208.9800    &  -0.460 &      &   kurucz        \\
NiI     &  4918.7060    &  -0.780 &	 &   Kurucz 	  & ZrII    &  4496.9600    &  -0.810 &      &   Bal81	       \\
NiI     &  4925.5590    &  -0.770 & D	 &   Kurucz 	  & 	    &		    &	      &     &	    	       \\ 
NiI     &  4935.8310    &  -0.350 & D	 &   FMW    	  & BaII    &	4554,0290   &  0,163  &   B  &   Kurucz        \\
NiI     &  4937.3410    &  -0.390 & D	 &   FMW    	  & BaII    &	4934,0760   &  -0,156 &   B  &   NBS           \\
NiI     &  4976.3260    &  -3.100 & C+	 &   NIST   	  & BaII    &	5853,6680   &  -1,510 &   B  &   NIST          \\
NiI     &  5003.7410    &  -2.800 & C+	 &   NIST   	  & BaII    &	6141,7130   &  -0,810 &   B  &   NIST         \\
NiI	&  5080.5280	&  0.330  &	 &   kurucz 	  & BaII    &	6496,8970   &  -1,010 &   B  &   NIST         \\    
NiI	&  5081.1070	&  0.300  &	 &   kurucz 	  & 	    &		    &	      &      &                \\     
NiI	&  5084.0890	&  0.090  &	 &   kurucz 	  & 	    &		    &	      &      &                \\    
NiI	&  5096.8540	&  -0.900 &	 &   kurucz 	  & 	     &  	     &         &      & 	  \\	
NiI	&  5099.9270	&  -0.100 &	 &   kurucz 	  & 	     &  	     &         &      & 	  \\	
NiI	&  5102.9660	&  -2.620 & C+   &   NIST   	  & 							      \\    
NiI	&  5137.0740	&  -1.990 & C+   &   NIST   	  & 							      \\    
NiI	&  5476.9040	&  -0.890 & C+   &   NIST   	  & 							      \\			  
NiI	&  5553.6900	&  -3.240 & C+   &   NIST   	  & 							      \\		  
NiI	&  5587.8580	&  -2.140 & C+   &   NIST   	  & 							      \\ 
NiI	&  5592.2620	&  -2.590 & C+   &   NIST   	  & 							      \\			  
NiI	&  5711.8880	&  -2.260 & C+   &   NIST   	  & 							      \\			
NiI	&  5754.6560	&  -2.340 & C+   &   NIST   	  & 							      \\			
NiI	&  5892.8720	&  -2.340 & C+   &   NIST   	  & 							      \\			
NiI	&  6163.4180	&  -0.682 &	 &   kurucz 	  & 							      \\		 
NiI	&  6170.5670	&  -1.808 &	 &   kurucz 	  &  								 \\
NiI	&  6175.3600	&  -0.530 &	 &   kurucz 	  &  								 \\
NiI	&  6176.8070	&  -0.260 &	 &   kurucz 	  &								 \\
        &     	        &    	  &	 &		  &								 \\ \hline

 \end{longtable}}     
\end{document}